\documentclass[aps,prd,12pt,twocolumn,superscriptaddress,showpacs,showkeys,reprint,longbibliography
]{revtex4-1}

\usepackage{amsmath,amsthm,latexsym,amssymb,amsfonts}
\usepackage{xcolor}
\usepackage[%
  colorlinks=true,
  urlcolor=blue,
  linkcolor=blue,
  citecolor=blue
]{hyperref}
\usepackage{breqn}

\makeatletter
\let\cat@comma@active\@empty
\makeatother

\usepackage{array}
\newcolumntype{L}[1]{>{\raggedright\let\newline\\\arraybackslash\hspace{0pt}}m{#1}}
\newcolumntype{C}[1]{>{\centering\let\newline\\\arraybackslash\hspace{0pt}}m{#1}}
\newcolumntype{R}[1]{>{\raggedleft\let\newline\\\arraybackslash\hspace{0pt}}m{#1}}
\usepackage{xparse}

\DeclareDocumentCommand{\Ag}{ s o }{ \IfBooleanTF{#1}
    { \IfValueTF{#2}{ \boldsymbol{\mathcal{A}}_{(#2)} }{ \boldsymbol{\mathcal{A}} } }
    { \IfValueTF{#2}{            {\mathcal{A}}_{(#2)} }{            {\mathcal{A}} } } }
\DeclareDocumentCommand{\Af}{ s o }{ \IfBooleanTF{#1}
    { \IfValueTF{#2}{ \boldsymbol{A}_{(#2)} }{ \boldsymbol{A} } }
    { \IfValueTF{#2}{            {A}_{(#2)} }{            {A} } } }

\newcommand{\cdf}[1][]{{\boldsymbol{\mathcal{D}}}{#1}\!}

\DeclareDocumentCommand{\Fg}{ s o }{ \IfBooleanTF{#1}
    { \IfValueTF{#2}{ \boldsymbol{\mathcal{F}}_{(#2)} }{ \boldsymbol{\mathcal{F}} } }
    { \IfValueTF{#2}{            {\mathcal{F}}_{(#2)} }{            {\mathcal{F}} } } }
\DeclareDocumentCommand{\Ff}{ s o }{ \IfBooleanTF{#1}
    { \IfValueTF{#2}{ \boldsymbol{F}_{(#2)} }{ \boldsymbol{F} } }
    { \IfValueTF{#2}{            {F}_{(#2)} }{            {F} } } }

\newcommand{\ga}{\gamma}

\newcommand{\Ga}{\Gamma}

\DeclareMathOperator{\st}{\star}

\newcommand{\Lap}{\nabla^2}

\newcommand{\Mi}{\mathcal{M}}

\newcommand{\R}{\mathbb{R}}

\DeclareDocumentCommand\Te{o o m }{\mathcal{T}{}^{#1}_{#2}(#3)}

\newcommand{\Z}{\mathbb{Z}}

\newcommand{\vph}{\ensuremath{\varphi}}

\newcommand{\Rif}[2]{\boldsymbol{\mathcal{R}}^{{#1}}{}_{{#2}}}

\newcommand{\Tr}{\operatorname{Tr}}

\newcommand{\beq}{\begin{equation}}
\newcommand{\eeq}{\end{equation}}
\newcommand{\ber}{\begin{eqnarray}}
\newcommand{\eer}{\end{eqnarray}}

\newcommand{\dn}[2]{{\mathrm{d}}^{#1}{#2}\;}

\newcommand*{\de}[1]{\mathop{\mathrm{d}#1}\nolimits}

\newcommand\UTFSM{Departamento de F\'\i sica, Universidad T\'{e}cnica Federico Santa Mar\'\i a, \\ Casilla 110-V, Valpara\'\i so, Chile}
\newcommand\CCTVal{Centro Cient\'\i fico Tecnol\'ogico de Valpara\'\i so, \\ Casilla 110-V, Valpara\'\i so, Chile}
\newcommand\CFF{Centro de F\'\i sica Fundamental,  Universidad de los Andes,\\ 5101 M\'erida, Venezuela}

\hypersetup{%
  pdftitle={Einstein's gravity from a polynomial affine model},
  pdfauthor={Oscar Castillo-Felisola,}{Aureliano Skirzewski},
  pdfkeywords={Affine Gravity,} {Torsion,} {Generalised Gravity.},
  pdflang={English}
}

\begin{document}

\title{Einstein's gravity from a polynomial affine model}

\author{Oscar \surname{Castillo-Felisola}}
\email{o.castillo.felisola@gmail.com}
\affiliation{\CCTVal.}
\affiliation{\UTFSM.}

\author{Aureliano \surname{Skirzewski}}
\email{askirz@gmail.com}
\affiliation{\CFF.}

\begin{abstract}
  We show that the effective field equations for a recently formulated polynomial affine model of gravity, in the sector of a torsion-free connection, accept general Einstein manifolds---with or without cosmological constant---as solutions. Moreover, the effective field equations are partially those obtained from a gravitational Yang--Mills theory known as Stephenson--Kilmister--Yang theory. Additionally, we find a generalization of a minimally coupled massless scalar field in General Relativity within a ``minimally'' coupled scalar field in this affine model. Finally, we present a brief analysis of the propagators of the gravitational theory, and count the degrees of freedom. For completeness we prove that a Birkhoff-like theorem is valid for the analyzed sector.
\end{abstract}


\maketitle

\section{\label{intro}Introduction}

A hundred years ago, A.~Einstein presented by the first time the correct field equations of its relativistic theory of gravitation, known as General Relativity~\cite{einstein1915feldgleichungen}. It was not only consistent with the available observations, but it also predicted several effects that have been consistently observed and measured---see Ref.~\cite{Will:2014kxa} for a recent review---, being the last one the confirmation of the existence of gravitational waves~\cite{Abbott:2016blz}. Thus, General Relativity has become the most successful theory of gravitational interactions nowadays.

However, General Relativity seems to be an effective theory of a, yet to be discovered, more fundamental description of gravitational effects. A first hint was pointed out by E.~Cartan, after remarking that the curvature is not built up from a metric field but from a connection, and that a part of it was left behind if one chooses to work with the Levi-Civita connection~\cite{Cartan1922,Cartan1923,Cartan1924,Cartan1925}. One could also see this by considering Palatini's approach in General Relativity, in which both metric and connection fields are independent, and noticing that generically the theory possesses terms which do not appear in General Relativity (when it is coupled to matter). In addition, we mention a few aspects related with the quantization: First, the fact that within the standard formalism of quantization, General Relativity is non-renormalizable~\cite{'tHooft:1973us,'tHooft:1974bx,Deser:1974cz,Deser:1974cy}, suggests the theory can be thought as an effective theory. Second, the Wheeler--DeWitt equations are not well-defined in general due to their nonpolynomial dependence on the Arnowitt--Deser--Misner (or ADM) variables~\cite{Arnowitt:1959ah,Arnowitt:1960es,WheelerGeo,DeWitt:1967yk,DeWitt:1967ub,DeWitt:1967uc}. One of the most successful quantization methods of gravitational interactions---developed from ideas by Gambini and Trias~\cite{Gambini:1980yz,Gambini:1986ew}, by Ashtekar~\cite{Ashtekar:1986yd,Ashtekar:1987gu}---, hints that a first order formulation of General Relativity is not as complete as expected, since the Lagrangian formulation should include an extra term known as Holst term~\cite{Holst:1995pc}. Finally, it is well known that once matter couples to gravity (in particular fermionic matter), the assumption of vanishing torsion in standard General Relativity is no longer preserved as a field equation, and the best approach is to consider Cartan's generalization of General Relativity~\cite{Kibble:1961ba,Hehl:1976kj,Shapiro:2001rz,Hammond:2002rm}.

Standing on the argument of the necessity of a quantum theory of gravity, several generalizations of Einstein's General Relativity have been proposed. One important class of these generalizations considers the connection field to be independent of the metric, known as Palatini type theories. Within this class of theories one finds the generalization due to Cartan---which considers a connection with nonvanishing torsion~\cite{Cartan1922,Cartan1923,Cartan1924,Cartan1925}---, but also a broader kind of generalization which do not require the metricity condition, see for example Refs.~\cite{Hehl:1976kj,Hehl:1994ue,Pagani:2015ema}.

Interestingly, gravitational theories with a generic linear connections have degrees of freedom which do not correspond to the expected massless graviton~\cite{Sezgin:1979zf}. Albeit these gravitational theories with an arbitrary connection possess ghosts, there are examples of well-behaved (stable) systems which have ghosts, such are the cases of the Pais--Uhlenbeck oscillator~\cite{Mannheim:2004qz,Bender:2007wu,Smilga:2008pr,Ilhan:2013xe}, or the higher derivative supersymmetric quantum mechanical system presented in Ref.~\cite{Robert:2008}. The last sentence, allows to argue that a model cannot be disregarded for the presence of ghosts model, without a rigorous analysis of its stability.

In order to undertake the premises of incompleteness and non-polynomial structure of the theory, in a previous paper~\cite{Skirzewski:2014eta}, we presented a polynomial (purely) affine gravitational model in four dimensions built up entirely on the basis of full diffeomorphism invariance. Worth to mention is that the most general action built this way contains only power-counting renormalizable terms (which by no means guarantees the renormalizability of the system). We showed that its non-relativistic limit, around a homogeneous and isotropic spacetime, yields a geodesic deviation corresponding to the generated by a Newtonian gravitational potential, despite the existence of torsion. We also highlighted the possibility of using standard methods to quantize the model, due to its polynomial structure (unlike the earliest proposals~\cite{Eddington1923math,schrodinger1950space}), and also the likelihood of avoiding the uniqueness of diffeomorphism-invariant states from Loop Quantum Gravity programme~\cite{Lewandowski:2005jk}, due to the absence of a fundamental metric, or in other words, the absence of flux operators.

Interestingly, by construction, the polynomial affine gravity proposed in Ref.~\cite{Skirzewski:2014eta} has no explicit terms leading to three-point graviton vertices, since all graviton \emph{self-interaction} is mediated by non-Riemannian parts of the connection. This feature might allow to bypass the general postulates supporting the no-go theorems stated in Refs.~\cite{McGady:2013sga,Camanho:2014apa}, where it was proven that generic three-point graviton interactions are highly constrained by causality and analyticity of the $S$-matrix, and the only \emph{acceptable} structure of the three-point graviton vertices is the one coming from General Relativity.

The aim of this work is to show that in the vanishing torsion sector, the field equations of the polynomial affine gravity are a generalization of the Einstein's equations. 
The paper is organized as follows. In Sec.~\ref{model}, we present briefly the polynomial affine gravity model, and summarize the results obtained in Ref.~\cite{Skirzewski:2014eta} concerning the Newtonian limit of the model. In Sec.~\ref{rlimit}, we restrict ourselves to the sector of the theory with vanishing torsion, and find the field equations; which surprisingly are a known generalization of the Einstein's field equation. Then, in Sec.~\ref{matter} we show that under certain considerations the coupling of the gravitational model with a scalar field represents a generalization of the minimally coupled system Einstein--Klein--Gordon. In Sec.~\ref{propag}, we present a brief analysis of the propagators of the model, which allows us to count the propagating degrees of freedom. In addition, we corroborate that result with a method equivalent to the one of Dirac~\cite{Diaz:2014yua}. Finally, in Sec.~\ref{conclusions} we summarize and discuss the possible implications of the results. In order to make the paper self-contained, we have included three appendices: In Appendix~\ref{DA}, we describe a dimensional analysis strategy that allowed us to build up the most general action consistent with diffeomorphisms invariance. In Appendix~\ref{Known}, we show a collection of known solutions to the effective field equations of the model. Finally, in Appendix~\ref{Schw}, we demonstrate---restricting ourselves to the vanishing torsion sector---the equivalent of the Birkhoff's theorem of the theory.

\section{\label{model}Polynomial affine gravity}

First of all, we highlight that the model constructed below has as only fundamental field an affine connection, and no metric field is needed neither for contracting nor lowering or raising indices. Moreover, in order to guarantee the correct transformation of the Lagrangian density, the geometrical objects used to write down the action will be the curvature and torsion of an affine connection, $\hat{\Gamma}^\mu{}_{\rho\sigma}$, which accepts a decomposition on irreducible components as
\begin{equation}
  \hat{\Gamma}^\mu{}_{\rho\sigma} = \hat{\Gamma}^\mu{}_{(\rho\sigma)} + \hat{\Gamma}^\mu{}_{[\rho\sigma]} = {\Gamma}^\mu{}_{\rho\sigma} + \epsilon_{\rho\sigma\lambda\kappa}T^{\mu,\lambda\kappa}+A_{[\rho}\delta^\mu_{\nu]},
\end{equation}
where ${\Gamma}^\mu{}_{\rho\sigma} = \hat{\Gamma}^\mu{}_{(\rho\sigma)}$ is symmetric in the lower indices, $A_\mu$ is a vector field corresponding to the trace of torsion, and  $T^{\mu,\lambda\kappa}$ is a Curtright field~\cite{Curtright:1980yk}, satisfying \mbox{$T^{\kappa,\mu\nu } = - T^{\kappa,\nu\mu }$} and $\epsilon_{\lambda\kappa\mu\nu}T^{\kappa,\mu\nu }=0$. Since no metric is present, the epsilon symbols are not related by raising (lowering) their indices, but instead we demand that \mbox{$\epsilon^{\delta\eta\lambda\kappa}\epsilon_{\mu\nu\rho\sigma}=4!\delta^{\delta}{}_{[\mu}\delta^\eta{}_{\nu}\delta^{\lambda}{}_{\rho} \delta^\kappa{}_{\sigma]}$.}
\begin{widetext}
Using the above decomposition, the most general action preserving diffeomorphisms is
  \begin{dmath}
    \label{4dfull}
    S[{\Gamma},T,A] =
    \int\dn{4}{x}\Bigg[
      B_1\, R_{\mu\nu}{}^{\mu}{}_{\rho} T^{\nu,\alpha\beta}T^{\rho,\gamma\delta}\epsilon_{\alpha\beta\gamma\delta}
      +B_2\, R_{\mu\nu}{}^{\sigma}{}_\rho T^{\beta,\mu\nu}T^{\rho,\gamma\delta}\epsilon_{\sigma\beta\gamma\delta}
      +B_3\, R_{\mu\nu}{}^{\mu}{}_{\rho} T^{\nu,\rho\sigma}A_\sigma
      +B_4\, R_{\mu\nu}{}^{\sigma}{}_\rho T^{\rho,\mu\nu}A_\sigma
      +B_5\, R_{\mu\nu}{}^{\rho}{}_\rho T^{\sigma,\mu\nu}A_\sigma
      +C_1\, R_{\mu\rho}{}^{\mu}{}_\nu \nabla_\sigma T^{\nu,\rho\sigma}
      +C_2\, R_{\mu\nu}{}^{\rho}{}_\rho \nabla_\sigma T^{\sigma,\mu\nu} 
      +D_1\, T^{\alpha,\mu\nu}T^{\beta,\rho\sigma}\nabla_\gamma T^{(\lambda, \kappa) \gamma}\epsilon_{\beta\mu\nu\lambda}\epsilon_{\alpha\rho\sigma\kappa}
      +D_2\,T^{\alpha,\mu\nu}T^{\lambda,\beta\gamma}\nabla_\lambda T^{\delta,\rho\sigma}\epsilon_{\alpha\beta\gamma\delta}\epsilon_{\mu\nu\rho\sigma}
      +D_3\,T^{\mu,\alpha\beta}T^{\lambda,\nu\gamma}\nabla_\lambda T^{\delta,\rho\sigma}\epsilon_{\alpha\beta\gamma\delta}\epsilon_{\mu\nu\rho\sigma}
      +D_4\,T^{\lambda,\mu\nu}T^{\kappa,\rho\sigma}\nabla_{(\lambda} A_{\kappa)} \epsilon_{\mu\nu\rho\sigma}
      +D_5\,T^{\lambda,\mu\nu}\nabla_{[\lambda}T^{\kappa,\rho\sigma} A_{\kappa]} \epsilon_{\mu\nu\rho\sigma}
      +D_6\,T^{\lambda,\mu\nu}A_\nu\nabla_{(\lambda} A_{\mu)}
      +D_7\,T^{\lambda,\mu\nu}A_\lambda\nabla_{[\mu} A_{\nu]} 
      +E_1\,\nabla_{(\rho} T^{\rho,\mu\nu}\nabla_{\sigma)} T^{\sigma,\lambda\kappa}\epsilon_{\mu\nu\lambda\kappa}
      +E_2\,\nabla_{(\lambda} T^{\lambda,\mu\nu}\nabla_{\mu)} A_\nu
      +T^{\alpha,\beta\gamma}T^{\delta,\eta\kappa}T^{\lambda,\mu\nu}T^{\rho,\sigma\tau}
      \Big(F_1\,\epsilon_{\beta\gamma\eta\kappa}\epsilon_{\alpha\rho\mu\nu}\epsilon_{\delta\lambda\sigma\tau}
      +F_2\,\epsilon_{\beta\lambda\eta\kappa}\epsilon_{\gamma\rho\mu\nu}\epsilon_{\alpha\delta\sigma\tau}\Big) 
      +F_3\, T^{\rho,\alpha\beta}T^{\gamma,\mu\nu}T^{\lambda,\sigma\tau}A_\tau \epsilon_{\alpha\beta\gamma\lambda}\epsilon_{\mu\nu\rho\sigma}
      +F_4\,T^{\eta,\alpha\beta}T^{\kappa,\gamma\delta}A_\eta A_\kappa\epsilon_{\alpha\beta\gamma\delta}\Bigg].
  \end{dmath}
\end{widetext}
In Eq.~\eqref{4dfull}, we have dropped all terms which can be related through partial integration, and those which are topological invariant (e.g. the Euler density). A straightforward dimensional analysis shows the impossibility of other contributions into the action, see Appendix~\ref{DA} (in particular Sec.~\ref{sec:ld}). Interestingly, the above action turns out to be power-counting renormalizable, which does not guarantee renormalizability, but is a nice feature. The structure of the model yields no three-point graviton vertices, which might allow to overcome  the \emph{no-go} theorems found in Refs.~\cite{McGady:2013sga,Camanho:2014apa}. It is worth mentioning that all coupling constants are dimensionless, which might be a hint of conformal invariance of the model~\cite{Buchholz:1976hz}, and further analysis of this invariance will be presented in Ref.~\cite{OCF-future3}.

Although there is no obvious equivalence between the action in Eq.~\eqref{4dfull} and General Relativity, particularly due to the lack of a source metric field, we analyzed in Ref.~\cite{Skirzewski:2014eta} the scalar perturbations and studied perturbative inhomogeneous sources to the connection field equations, by considering a generic matter action. In the rest of this section we summarize the procedure which leads to the Newtonian limit of the model.

First, we considered a static, homogeneous, and isotropic expansion of fields,
\begin{align}
  A_\mu &= \delta_\mu^0 A + a_\mu, \notag \\
  T^{\mu,\nu\rho} &= \delta^{\mu}_m\delta^{\nu\rho}_{m0}T + t^{\mu,\nu\rho},
  \label{GammaExp} \\
  \Gamma^\lambda{}_{\mu\nu} &= E \delta^\lambda_0 \delta^m_\mu \delta^m_\nu + F \delta^\lambda_m \delta^m_{(\mu}\delta^0_{\nu)} + G\delta^\lambda_0 \delta^0_{\mu}\delta^0_{\nu} + \gamma^\lambda{}_{\mu\nu}, \notag
\end{align}
with $\delta^{\mu\nu}_{\lambda\kappa}=\delta^{\mu}_{\lambda}\delta^{\nu}_{\kappa}-\delta^{\mu}_{\kappa}\delta^{\nu}_{\lambda}$. Next, by analysing the first order perturbations of the actions, we found a choice of the coupling constants which allows nontrivial solutions of these field equations, together with the non-relativistic limit of the geodesic equation. Then, we substitute the field components by their scalar perturbation decomposition,
\begin{equation}
  a_\mu \to \delta_\mu^0 a+\delta_\mu^m \partial_{m}a_{\mathtt{L}},
\end{equation}
\begin{equation}
  \begin{split}
    t^{\mu,\nu\rho} &\to \delta^{\mu}_m\delta^{\nu\rho}_{n0} \Big(t \delta^{m n} + \partial^m \partial^n t_{\mathtt{L}} \Big)
    +\delta^{\mu}_0 \delta^{\nu\rho}_{m0} \partial^m c_{\mathtt{L}}
    \\
    & \quad + \Big(\delta^{\mu}_0\delta^{\nu\rho}_{mn}-\delta^{\mu}_m\delta^{\nu\rho}_{n0}\Big)\epsilon^{m n p} \partial_{p} b
    \\
    & \quad +\delta^{\mu}_m \delta^{\nu}_{n} \delta^{\rho}_{p} \Big(\epsilon^{n p q}\partial_q \partial^m d_1 +  (\delta^{m n} \partial^p - \delta^{m p} \partial^n)d_2\Big)
  \end{split}
\end{equation}
and
\begin{equation}
  \begin{split}
    \gamma^\lambda_{\mu\nu}
    &\to
    \delta^\lambda_0\delta^0_\mu\delta^0_\nu u 
    + \delta^\lambda_m \delta^0_\mu\delta^0_\nu \partial^m v_{\mathtt{L}}
    + 2\delta^\lambda_0 \delta^0_{(\mu}\delta^m_{\nu)} \partial_m w_{\mathtt{L}}
    \\
    & \quad + \delta^\lambda_0 \delta^m_\mu\delta^n_\nu \Big(x \delta_{mn} + \partial_m \partial_n x_{\mathtt{L}}\Big)
    \\
    & \quad + 2\delta^\lambda_m \delta^0_{(\mu}\delta^n_{\nu)} \Big(y_1 \delta^m{}_n + \epsilon^{m p}{}_{n} \partial_p y_2 + \partial^m \partial_n y_{\mathtt{L}}\Big)
    \\
    & \quad + \delta^\lambda_m \delta^n_{\mu}\delta^p_{\nu} \Big(\delta_{n p} \partial^m z_1 + (\delta^m{}_n \partial_p+\delta^m{}_p \partial_n) z_2
    \\
    & \qquad +  (\epsilon^{m q}{}_n \partial_p+\epsilon^{m q}{}_p \partial_n) \partial_q z_3 + \partial^m \partial_n \partial_p z_{\mathtt{L}}\Big),
  \end{split}
\end{equation}
where the ``\texttt{L}'' sub-index identifies the longitudinal degrees of freedom.


Albeit the inclusion of matter in a non-metric spacetime is a nonstandard procedure, we assumed that the matter action would depend on the most general, symmetric $\binom{2}{0}$-tensor density, $\mathfrak{g}^{\mu \nu}$, built up with the available fields. The dimensional analysis in Sec.~\ref{sec:im} shows that this tensor density contains the same terms as what we call the ``Eddington's metric density''~\footnote{Despite the suggestive name, this is a $\binom{2}{0}$-tensor density which does not necessarily satisfy the metric conditions. However, for the simple case of the Einstein--Hilbert action, it is in fact the inverse metric density.}, defined as the variation of the action with respect to the symmetric part of the Ricci tensor~(see Refs.~\cite{Eddington1923math,schrodinger1950space,Poplawski:2012bw}), i.e.,
\begin{equation}
  \label{metric}
  \frac{\delta\ }{\delta R_{(\mu\nu)}} S[\Gamma] \equiv \sqrt{\mathsf{g}} \, \mathsf{g}^{\mu\nu} = \bar{g}^{\mu\nu}.
\end{equation}

From the geodesic equations and the linearized action on scalar perturbation, we found that only a few terms in the scalar perturbation affect the geodesic equation~\footnote{We cross-check our result using the software Cadabra~\cite{Peeters2007550,peeters2007symbolic,Peeters:2007wn}.}. The effective geodesic deviation is given by
\begin{equation}
  \label{NewtonPot}
  \Gamma^i{}_{00} = - \frac{1}{8\pi} \frac{ \partial\mathcal{L}_{\text{Matter}} }{ \partial \mathfrak{g}^{00} } \, \nabla^i \left(\frac{1}{|\vec{x}|}\right),
\end{equation}
is the usual Newtonian force induced by a massive source, where we restricted ourselves to consider $\mathfrak{g}^{\mu \nu} = \bar{g}^{\mu\nu}$.

\section{\label{rlimit}Relativistic limit}
Despite we obtained the correct Newtonian gravitational potential in the non-relativistic limit, see Eq.~\eqref{NewtonPot}, a \emph{post}-Newtonian approximation is necessary to explain gravitational phenomena like Mercury's perihelion and the bending of light by a gravitational source. Therefore, we should go beyond the Newtonian limit in order to compare our model with General Relativity.

Even though a comparison between the models does not require the imposition of vanishing torsion, for the sake of simplicity, at this stage we shall focus on a sector of the theory in which the connection is torsion-free. Notice that the vanishing torsion limit---equivalent to take $T^{\lambda,\mu\nu} \to 0$ and $A_\mu \to 0$---cannot be taken at the action level, but in the field equations. Interestingly, the field equations from Eq.~\eqref{4dfull} can be consistently \emph{truncated} under such requirements. It can be realized as follows, since only the terms of the action linear in these fields will be relevant, one can consider the effective action linear in the torsion's fields, i.e.,
\begin{equation}
  \label{eff-action}
  S_{\text{eff}} = \int\dn{4}{x} \Big( C_1\, R_{\lambda\mu}{}^{\lambda}{}_\nu \nabla_\rho 
  + C_2 \, R_{\mu\rho}{}^{\lambda}{}_\lambda \nabla_\nu \Big) T^{\nu,\mu\rho} ,
\end{equation}
and the only nontrivial field equation after the limit will be the one for the Curtright, $T^{\nu,\mu\rho}$,
\begin{equation}
  \nabla_{[\rho} R_{\mu]\nu} + \kappa \nabla_{\nu} R_{\mu\rho}{}^\lambda{}_\lambda = 0,
  \label{almostSimpleEOM}
\end{equation}
with $\kappa$ a constant related with the original couplings of the model. We assume that the connection is compatible with volume form, i.e., it is equi-affine~\cite{nomizu1994affine,MO-Bryant02}.  The equi-affine condition assures that the Ricci tensor of the connection is symmetric, and the contraction of the last indices vanishes, thus the second term in the field equation is absent and the gravitational equations are
\begin{equation}
  \nabla_{[\rho} R_{\mu]\nu} = 0.
  \label{SimpleEOM}
\end{equation}

Equation~\eqref{SimpleEOM} is a generalization of a condition known in Riemannian geometry as covariantly constant Ricci curvature---aka parallel Ricci curvature---, \mbox{$\nabla_{\rho} R_{\mu\nu} = 0$.}  All Einstein manifolds, whose Ricci tensor is proportional to the metric, \mbox{$R_{\mu\nu} \propto g_{\mu\nu}$,} satisfy the parallel Ricci condition due to the metricity condition, and therefore every vacuum solution to the Einstein's equations solves the (simplified) field equations of our model. Consequently, the fact that the non-relativistic limit of the gravitational potential in Eq.~\eqref{NewtonPot} yields a Newtonian potential seems clearer, and we can argue that even the post-Newtonian corrections coming from General Relativity are present in the chosen scenario of our model. An additional comment, the parallel Ricci curvature assures that although the manifold is not Einstein in general, the metric has to be locally a product of Einstein metrics~\cite{Besse}.



The Eq.~\eqref{SimpleEOM} is related through the second Bianchi identity to the harmonic curvature condition~\cite{bourguignon1981varietes},
\begin{equation}
  \label{harm-curv}
  \nabla_\lambda R_{\mu\nu}{}^\lambda{}_\rho = 0,
\end{equation}
and it can be shown that a manifold with harmonic curvature is equivalent (in four dimensions) to a manifold with harmonic Weyl tensor and constant scalar curvature~\cite{Berger:1969}, or in other words the Ricci tensor is a Codazzi tensor~\footnote{A Codazzi tensor is a symmetric $(0,2)$-type tensor, $T$, satisfying the condition $D_X T(Y,Z) = D_Y T(X,Z)$~\cite{Derdzinski01071983}.}. For proofs of these equivalences, see Refs.~\cite{Derdzinski:1985,Besse}.

Notice that Eqs.~\eqref{SimpleEOM} and~\eqref{harm-curv} accept a geometrical interpretation equivalent to that of the field equations of a pure Yang--Mills theory, which in the language of differential forms are
\begin{align}
  \cdf \Ff* &= 0, & \cdf \st \Ff* &= 0,
\end{align}
where $\Ff* = \cdf \Af*$ is the field strength 2-form (the curvature 2-form of the connection in the principal bundle, see for example Ref.~\cite{bourguignon1982yang,Nakahara}), and the operator $\st$ denotes the Hodge star. Now, these Yang--Mills field equations are obtained from the variation of the action functional
\begin{equation}
  S_{\textsc{ym}} = \int \Tr \Big( \Ff* \st \Ff* \Big),
\end{equation}
and the Jacobi identity for the covariant derivative.

Interestingly, the Eq.~\eqref{SimpleEOM}---equivalently Eq.~\eqref{harm-curv}---can be obtained from an effective gravitational Yang--Mills functional action~\cite{stephenson1958quadratic,kilmister1961use,Yang1974}, 
\begin{equation}
  \label{SKY}
  S_{\textsc{ym}} = \int \Tr \left( \Rif{}{} \st \Rif{}{} \right) = \int \left( \Rif{a}{b} \st \Rif{b}{a} \right),
\end{equation}
where $\Rif{}{} \in \Omega^2(\Mi, T^*\Mi \otimes T\Mi)$ is the curvature two-form, the operator $\st$ denotes the Hodge star, and the trace is taken on the bundle indices (see Ref.~\cite{bourguignon1982yang}).

The gravitational model described by the action in Eq.~\eqref{SKY} is called Stephenson--Kilmister--Yang (or SKY for short). However, the standard interpretation of this model requires a metric tensor. A quick analysis of the field equations \`a la Palatini shows that Eqs.~\eqref{SimpleEOM} and~\eqref{harm-curv} are obtained only from the variation with respect to the connection. Albeit this effective theory has been widely studied, the extra field equation (for the metric field) strongly constraints the solutions of the gravitational Yang--Mills~\cite{JZcomm}. However, that complication does not affect our affine interpretation of the theory, where the metric is absent.

Intriguingly enough, the effective theory is not renormalizable---according to the arguments in Refs.~\cite{McGady:2013sga,Camanho:2014apa}---, although it has been discussed in Ref.~\cite{Chen:2010at} that at least the ghost problem can be avoided through the arguments exposed in Refs.~\cite{Kleinert:1987eb,Bender:2007wu,Bender:2008vh,Mannheim:2009zj}.

\section{\label{matter}Polynomial affine gravity coupled with a scalar field}

Until now, the most general diffeomorphism-invariant and power-counting renormalizable (gravitational) theory for an affine connection has been built, and we have showed that in a certain sector it is a generalization of General Relativity. In what follows, we show an attempt of including scalar matter into the model. The theory does not require a spacetime metric, but instead we should consider the most general, symmetric $\binom{2}{0}$-tensor density, $\mathfrak{g}^{\mu\nu}$,  built with the available fields (see Appendix~\ref{sec:im}), and use it to build Lagrangian densities for the matter content. Following our precept, the matter content---scalar matter---should couple to $\mathfrak{g}^{\mu\nu}$, given by
\begin{dmath}
  \mathfrak{g}^{\mu\nu} = \alpha \, \nabla_\lambda T^{\mu,\nu\lambda} + \beta \, A_\lambda T^{\mu,\nu\lambda} + \gamma \, \epsilon_{\lambda\kappa\rho\sigma} T^{\mu, \lambda\kappa} T^{\nu, \rho\sigma},
  \label{geng}
\end{dmath}
with $\alpha$, $\beta$ and $\gamma$ arbitrary coefficients.

We consider the action provided by the ``kinetic term''
\begin{dmath}
  \label{ScalarAction}
  S_\phi = -  \int \dn{4}{x} \Big( \alpha \, \nabla_\lambda T^{\mu,\nu\lambda}  + \beta \, A_\lambda T^{\mu,\nu\lambda} + \gamma \, \epsilon_{\lambda\kappa\rho\sigma} T^{\mu, \lambda\kappa} T^{\nu, \rho\sigma} \Big) \partial_\mu\phi\partial_\nu\phi,
\end{dmath}
which makes a nontrivial contribution to the field equations once we restrict to the sector of interest.

The equation for the Curtright field when the scalar field is turned on is (without lost of generality we fixed the coefficient $C_1 = 1$)
\begin{equation*}
  \nabla_{[\sigma} R_{\rho]\mu}{}^{\mu}{}_\nu - {C_2} \nabla_\nu  R_{\rho\sigma}{}^{\mu}{}_\mu - \alpha \nabla_{[\sigma} \Big( \partial_{\rho]}\phi \partial_\nu\phi \Big) = 0,
\end{equation*}
which under our considerations simplifies to 
\begin{equation}
  \nabla_{[\sigma} R_{\rho]\nu} - \alpha \nabla_{[\sigma} \Big( \partial_{\rho]}\phi \partial_\nu\phi \Big) = 0.
  \label{SimpleEOMwS}
\end{equation}
In that case, we find a particular solution of Eq.~\eqref{SimpleEOMwS},
\begin{equation*}
  R_{\mu\nu} - \alpha \partial_{\mu} \phi \partial_{\nu} \phi = \Lambda g_{\mu\nu},
\end{equation*}
where we have used a suggestive notation by denoting with $g_{\mu\nu}$ a covariantly constant, invertible, and symmetric two-tensor. Notice that $\alpha$ is the gravitational coupling constant, which in Einstein--Hilbert gravity is proportional to $G_N$.
This equation can be written in the more conventional form
\begin{equation}
  R_{\mu\nu} - \frac{1}{2} g_{\mu\nu} R + \Lambda g_{\mu\nu} = \alpha \Big( \partial_{\mu} \phi \partial_{\nu} \phi - \frac{1}{2} g_{\mu\nu} \big( \partial\phi \big)^2 \Big).
\end{equation}
Additionally, the second Bianchi identity imposes
\begin{equation}
  \nabla^\mu \partial_{\mu} \phi = 0.
\end{equation}
This condition is, in the sense argued in Ref.~\cite{Bekenstein:2014uwa}, the equation of motion for the scalar field. Notice that, the Euler--Lagrange equation of motion for the scalar field yields no information after taking the vanishing torsion \emph{truncation}.

It is well-known that this system of equations can be obtained effectively from the Einstein--Hilbert action coupled minimally to a massless scalar field
\begin{equation}
  S_{\text{eff}} = \frac{1}{\alpha} \int \dn{4}{x} \sqrt{g} \left( R + 2 \Lambda - \frac{\alpha}{2} g^{\mu\nu} \partial_{\mu}\phi \partial_{\nu} \phi \right).
\end{equation}
We have show, so far, that our model of polynomial purely affine gravity is compatible with General Relativity in the sector of vanishing torsion, and in addition the compatibility can be extended to include the interaction with a free scalar field. The analysis of this coupled system could be enhanced to include a potential for the scalar field, however, we are not considering this case at the moment.

\section{\label{propag}Analysis of propagators}

A step toward the analysis of renormalizability of the model should be the computation of the free propagators of the model around a given background. The choice of background is necessary in order to simplify the computations.  Our first task is to expand up to second order in the fields around a maximally symmetric Riemannian background. Maximally symmetric spaces possess remarkably simple curvature tensors~\cite{weinberg1972gravitation},
\begin{equation}
  R_{\mu\nu}{}^\lambda{}_\rho = {K} \Big( g_{\nu\rho} \delta^\lambda_\mu - g_{\mu\rho} \delta^\lambda_\nu \Big),
\end{equation}
and the Einstein's field equations in vacuum
\begin{equation}
  R_{\mu\nu} = \Lambda \, g_{\mu\nu},
\end{equation}
require that $K = \frac{\Lambda}{3}$. 
Therefore, in notation of Eqs.~\eqref{GammaExp} we have that $A = T = 0$, and the contribution due to the torsion and non-metricity enters as a perturbative effect. Thus, most terms in the action~\eqref{4dfull} do not contribute to the propagator analysis. Additionally, the first order perturbation of the curvature tensor is
\begin{equation}
  R_{\mu\nu}{}^\lambda{}_\rho = \frac{\Lambda}{3}\,   {\delta}^{\lambda}_{\mu} {g}_{\rho \nu} - \frac{\Lambda}{3}\,   {\delta}^{\lambda}_{\nu} {g}_{\rho \mu} + {\nabla}_{\mu}{{\gamma}^{\lambda}\,_{\rho \nu}}\,  - {\nabla}_{\nu}{{\gamma}^{\lambda}\,_{\rho \mu}}.
\end{equation}
Under these assumptions, the relevant terms of the action~\eqref{4dfull} are those accompanied by the couplings $B_i$, $C_i$ and $E_i$, and the second order perturbation of the action yields
\begin{widetext}
  \begin{dmath}[compact, spread=2pt]
    \label{4dflatx}
    S^{(2)}[\gamma,t,a] =
    \int\dn{4}{x} \bigg[
      C_1 \big( - \Lambda {\gamma}^{\alpha}\,_{\beta \delta} {g}_{\alpha \gamma} {t}^{\beta \delta \gamma} + {\nabla}_{\alpha}{{\gamma}^{\alpha}\,_{\beta \delta}}\,  {\nabla}_{\gamma}{{t}^{\beta \delta \gamma}}\,  - {\nabla}_{\alpha}{{\gamma}^{\beta}\,_{\beta \delta}}\,  {\nabla}_{\gamma}{{t}^{\delta \alpha \gamma}}\, \big)
      + 2\, C_2 {\nabla}_{\alpha}{{\gamma}^{\beta}\,_{\beta \delta}}\,  {\nabla}_{\gamma}{{t}^{\gamma \alpha \delta}}\,
      + E_1 {\epsilon}_{\alpha \beta \delta \eta} {\nabla}_{\gamma}{{t}^{\gamma \alpha \beta}}\,  {\nabla}_{\lambda}{{t}^{\lambda \delta \eta}}\,
      + E_2 \big({\nabla}_{\alpha}{{a}_{\beta}}\,  {\nabla}_{\gamma}{{t}^{\gamma \alpha \beta}}\,  - {\nabla}_{\alpha}{{a}_{\beta}}\,  {\nabla}_{\gamma}{{t}^{\alpha \beta \gamma}}\, \big)
      + \Big(B_1 + \frac{1}{3}\, B_2 \Big) \Lambda {\epsilon}_{\alpha \beta \gamma \delta} {g}_{\eta \lambda} {t}^{\eta \alpha \beta} {t}^{\lambda \gamma \delta}
      + \Big( - B_3 + \frac{2}{3}\, B_4 \Big) \Lambda {a}_{\alpha} {g}_{\beta \gamma} {t}^{\beta \alpha \gamma}
      \bigg].
    \label{4dmax}
  \end{dmath}
\end{widetext}
Longitudinal and transverse projectors can be introduced in order to decompose the action and specify in a more standard fashion the dynamics of Eq.~\eqref{4dmax}. This can be done since we have the metric $g_{\mu\nu}$ compatible with the connection, and $\nabla^\mu=g^{\mu\nu}\nabla_\nu$. Additionally we will denote by $\Delta = \nabla^\mu\nabla_\mu$ the covariant Laplacian. Let us define $P_{\mathtt{L}}{}^\mu{}_\nu=\nabla^\mu\Delta^{-1} \nabla_\nu $, the projector into logitudinal vectors, and similarly $P_\perp{}^\mu{}_\nu=\delta^\mu_\nu -P_{\mathtt{L}}{}^\mu{}_\nu$ is the projector into transverse vectors. Therefore, transverse indices are defined such that $\nabla_\mu A_\perp{}^{\mu}=0$.

Thus, the perturbations can be split into their longitudinal and transverse components as
\begin{dmath}
  a_\mu = {a_\perp}_{\mu} + {\nabla}_{\mu}{a_{\mathtt{L}}},
\end{dmath}
\begin{dmath}
  {t}^{\lambda \mu \nu} =
  {t_\perp}^{\lambda \mu \nu}
  + 2  {\nabla}^{\lambda} {t_{\mathtt{a}}^{\mu \nu}}
  - 2  {\nabla}^{[\mu}{{t_{\mathtt{a}}}^{\nu] \lambda}} 
  + 2  {\nabla}^{\mu}{{t_{\mathtt{s}}}^{\nu \lambda}}
  + 2  {\nabla}^{\lambda [\mu}{{t_{\mathtt{v}}}^{\nu]}}
\end{dmath}
and
\begin{dmath}
  {\gamma}^{\lambda}\,_{\mu \nu}  =
  {\gamma_\perp}^{\lambda}{}_{\mu \nu}
  + {\nabla}^{\lambda}{{\gamma_{\mathtt{s}}}_{\mu \nu}}
  + 2  {\nabla}_{(\mu}{{\gamma}^{\lambda}\,_{\nu)}}
  + 2  {\nabla}^{\lambda}\nabla_{(\mu}{{\gamma_{\mathtt{v}}}_{\nu)}}
  + 2  \nabla_{(\mu} \nabla_{\nu)} {{\gamma_{\mathtt{v}'}}^{\lambda}} 
  + 2  {\nabla}^{\lambda}\nabla_{(\mu} \nabla_{\nu)}{\gamma_{\mathtt{L}}},
\end{dmath}
where all the indices are transverse except those of the derivative $\nabla_\mu$, the subindex $\perp$ indicates that it corresponds to the fully transverse projection of the tensor in all of its indices, the subindex ${\mathtt{L}}$ indicates that this is the longitudinal projection of the tensor, while the subindices $\mathtt{s}$, $\mathtt{a}$ and $\mathtt{v}$ specify whether the tensor is symmetric, antisymmetric or a vector representation.

\begin{widetext}
  The perturbed action is then
  \begin{dmath}[compact, spread=2pt]
  \label{pert-Lambda}
    S = \int \dn{4}{x} \Bigg(
    C_1 \Big( {t_{\mathtt{s}}}^{\alpha\beta }-\frac{1}{4}g^{\alpha\beta}{{t_{\mathtt{s}}}^{\gamma}{}_{\gamma}} \Big) \bigg( -  (\frac{4\Lambda^2}{3}+\Lambda \Delta+\Delta^2) {\gamma_{\mathtt{s}}}_{\alpha\beta}  
    - \frac{8}{3}   \Lambda\Delta {{\gamma}_{\alpha\beta}} 
    \bigg)
    + {{t_{\mathtt{v}}}^{\alpha}}\bigg(
    \Big(( C_1 - 2\, C_2) \Delta^3 + \frac{1}{3}( C_1 + 2C_2 )\Lambda\Delta^2 + \frac{2}{3}( 2C_1 +  C_2 )\Lambda^2\Delta - \frac{1}{3}C_1  \Lambda^3\Big){{\gamma_{\mathtt{v}}}_{\alpha}}
    + \Big((C_1 - 2\, C_2) \Lambda\Delta - 2\, C_2 \Delta^2\Big){{\gamma_\perp}^{\beta}\,_{\beta \alpha}}
    + 8\, E_1 {\epsilon}_{\alpha \beta \gamma \delta} \Delta{\nabla}^{\beta }\Delta{{t_{\mathtt{a}}}^{\gamma\delta}}  
    +\Big(- 2\, E_2 \Delta^2 +  (- E_2 + B_3 - \frac{2}{3}\, B_4) \Lambda\Delta\Big){a_\perp}_{\alpha} 
    \bigg)
    + {{t_{\mathtt{s}}}^{\alpha}{}_{\alpha}}\bigg( \frac{2}{3}C_1 \Lambda\Delta{{\gamma}^{\beta}\,_{\beta}} 
    + \frac{2}{3}E_2 \Lambda \Delta{a_{\mathtt{L}}}  
    - \frac{4}{3}C_1 \Lambda^2 \Delta{g_{\mathtt{L}}}
    - \frac{1}{4}C_1 (\Lambda + \Delta)  \Delta{{\gamma_{\mathtt{s}}}^{\beta}{}_{\beta}}
    \bigg)   \Bigg),
  \end{dmath}
\end{widetext}
where $\Delta \equiv \nabla^{\alpha}\nabla_{\alpha}$ is the Laplacian operator. This action sets the dynamics of two transverse-traceless tensors (or specific combination of tensors), as well as two transverse vectors and two scalars, which are associated to eighteen degrees of freedom.

We have checked, using the method introduced in Ref.~\cite{Diaz:2014yua}, that the formal counting of degrees of freedom for the effective action in Eq.~\eqref{eff-action} yields eighteen degrees of freedom. We discover during the implementation of these program, that our model presents an extra symmetry associated to the metric independence property. 

In addition, when we restrict ourselves to perturbations around a flat (Minkowski) spacetime, the above equation simplifies greatly,
\begin{dmath}
  S = \int \dn{4}{x} \, \bigg[
    - C_1 ( \Delta\gamma_{\mathtt{s}\mu\nu})^* \Delta t_{\mathtt{s}}^{\mu\nu}
    + \Big( ( C_1 - 2 C_2) \nabla^\nu \Delta \nabla_\nu \gamma_{\mathtt{v}\mu}
    - 2 C_2 \Delta \gamma^\nu_{\perp\nu\mu}
    - 8 E_1 \epsilon_{\alpha\beta\nu\mu} \nabla^{\alpha} \Delta t_{\mathtt{a}}^{\beta\nu}
    - 2 E_2 \Delta a_{\perp\mu} \Big)^* \Delta t_{\mathtt{v}}^\mu
    \bigg],
\end{dmath}
and it is clearer that there are only eighteen degrees of freedom.

\section{\label{conclusions}Conclusions}

In this paper we have shown that the \emph{polynomial affine gravity} presented in Ref.~\cite{Skirzewski:2014eta} is the most general (polynomial) model for an affine theory of gravity, see the Appendix~\ref{sec:ld}, consistent with invariance under diffeomorphisms. The action in Eq.~\eqref{4dfull}, is parametrized by a nineteen-dimensional moduli space---after the boundary and topological terms have been dropped---, which in some sense made us wonder about a possible \emph{duality} between this model and the one proposed in Ref.~\cite{Pagani:2015ema}.

The field equations obtained from the action~\eqref{4dfull} are a set of very complex coupled partial differential equations. However, the system is consistent with the limit of $T^{\lambda,\mu\nu}$ and $A_\mu$ going to zero. This consistency is what is usually called a \emph{consistent truncation} of the theory. In this paper we have restricted ourselves to such a truncation of our model, which constrained us to the torsion-less sector of the moduli space. Additionally, this truncation leaves only one nontrivial field equation, Eq.~\eqref{almostSimpleEOM}.
Moreover, by considering torsion-free equi-affine connections---those preserving a volume element---or a Levi-Civita connection---in the cases where the manifold is metric---, the remaining field equation simplifies even further to Eq.~\eqref{SimpleEOM}.

The equation~\eqref{SimpleEOM} is a generalization of the field equations in General Relativity, in the sense that, any Einstein manifold (whose Ricci tensor is proportional to the metric) satisfy the parallel Ricci condition,
\begin{equation}
  \nabla_{\lambda} R_{\mu\nu} = 0.
\end{equation}
Although all Einstein manifolds satisfy the parallel Ricci condition, the contrary does not hold. Therefore, in Appendix~\ref{Known} we present a summary of known (metric) solutions to Eq.~\eqref{SimpleEOM}. In addition, in Appendix~\ref{Schw} we prove that a metric ansatz with spherical symmetry solves the field equation only if it is static, i.e., a Birkhoff-like theorem for our model. However, the non-linearity of the equations does not assure the uniqueness of the solution. Nonetheless, certain equations become linear if the functions $A$ and $B$ in Eq.~\eqref{ABmetric} are equal. Therefore, the unique solution is the (Anti-)de-Sitter--Schwarzschild solution.

We also pointed that our field equation~\eqref{SimpleEOM} can be obtained from an effective action, known as Stephenson--Kilmister--Yang (SKY) theory~\cite{stephenson1958quadratic,kilmister1961use,Yang1974}, which is built up in the spirit of a Yang--Mills theory for gravity. Despite the SKY theory has been criticized by many people, including C.~N.~Yang himself~\cite{JZcomm}, because its field equations are higher order differential equations for the metric field,
in our model the fundamental gravitational field is the connection. Equation~\eqref{SimpleEOM} is a second order partial differential equation for the connection. Although we have presented a series of metric solutions of the effective field equations (restricted to the torsion-free sector of the theory), the correct way to proceed is to propose a connection ansatz compatible with the symmetries of the problem. Using this procedure, we expect to find a non-metric solution to the field equation~\cite{OCF-future2}.

We would like to highlight that, since all gravitational effects described by standard General Relativity are contained in our model, particularly in the sector of metric compatible, torsion-free connections, there is enough liberty to start considering new cosmological effects coming from both nonvanishing torsion and non-metric connections with vanishing torsion, beside the results reported in Ref.~\cite{Chen:2013kia,Chen:2013ota}.

As a first step toward the coupling of matter with our polynomial affine gravity, we showed that there is a construct which might play the role of an inverse metric density. Notice that such a tensor density \emph{does not} necessarily satisfy a non-degeneracy condition as required for a metric, but even though it is an interesting question to ask under what considerations does this construct satisfy the properties of a metric. Using this tensor density, we proposed a \emph{kinetic} term for a scalar field (under diffeomorphisms), and show that the effective field equations are a generalization to those of General Relativity, when we restrict ourselves to the torsion-free sector. Despite the fact that in this sector the field equation for the scalar field collapses, and cannot be obtained from the variation of the action, using the algorithm proposed in Ref.~\cite{Bekenstein:2014uwa}, the usual scalar field equation is recovered in virtue of the second Bianchi identity. 

Finally, we analyzed the second order perturbation of the action in Eq.~\eqref{4dfull}, around a maximally symmetric spacetime. From this analysis it is possible to conclude that, at perturbative level, the number of degrees of freedom is eighteen, coming from a pair of transverse, symmetric, rank-two-tensors, and a pair of transverse vectors. Moreover, we applied the Lagrangian equivalent of Dirac's formalism for constrained systems (see Ref.~\cite{Diaz:2014yua}), and obtain the same number of physical states---it is a nice feature, because there are known physical systems for which the perturbative analysis of degrees of freedom do not coincide with the exact one~\footnote{We want to thank to M.~Blagojevi\'c, B.~Cvetkovi\'c and O.~Mi\v{s}kovi\'c for clarifications in this respect.}.


\begin{acknowledgments}
  We thank to N.~Pantoja, A.~Melfo, M.~Blagojevi\'c, B.~Cvetkovi\'c, O.~Mi\v{s}kovi\'c, C.~Corral, V.~Sharma, O.~Orellana and R.~L.~Bryant for their helpful discussions and inspiring comments, to J.~Zanelli for his suggestions on the physical insight into the problem and careful but critical review of the manuscript.
  This work was partially supported by CONICYT (Chile) under project No. 79140040. The ``Centro Cient\'ifico y Tecnol\'ogico de Valpara\'iso'' \mbox{(CCTVal)} is funded by the
Chilean Government through the Centers of Excellence Base Financing Program of CONICYT.
\end{acknowledgments}

\appendix

\section{\label{DA}Dimensional analysis method}

We built the model using six ingredients, a Curtright ($T^{\mu,\nu\lambda}$), a vector ($A_\mu$), the covariant derivative defined with the Levi-Civita connection ($\nabla_{\mu}$), both Levi-Civita tensors ($\epsilon_{\mu\nu\lambda\rho}$ and $\epsilon^{\mu\nu\lambda\rho}$), and the Riemannian curvature ($R_{\mu\nu}{}^\lambda{}_\rho$).
Since the Riemannian curvature is defined as the commutator of the covariant derivative, it is not an independent field, so it will be left out of the analysis, and only five ingredients remain.

Our interest is in general to build tensor densities. Therefore, we need to account for the number of \emph{free} indices, and weight density of these quantities. Denote by $N(\Phi)$  the operator which count the number (and position) of indices of the field $\Phi$, being positive (negative) for upper (lower) indices. Thus, we have that
\begin{equation*}
  \begin{aligned}
    N( T^{\mu,\nu\lambda} ) &= 3 & N( A_\mu ) &= -1 & N(  \nabla_{\mu} ) &= -1 \\
    N( \epsilon_{\mu\nu\lambda\rho} ) &= -4 & & & N( \epsilon^{\mu\nu\lambda\rho} ) &= 4 
  \end{aligned}
\end{equation*}
Then, for a general expression of the form
\begin{equation*}
  T^a A^b \nabla^c {\epsilon_{\dots}}^d {\epsilon^{\dots}}^e,
\end{equation*}
the indices counting yield
\begin{equation}
  N( T^a A^b \nabla^c {\epsilon_{\dots}}^d {\epsilon^{\dots}}^e ) = n,
\end{equation}
with
\begin{equation}
  n = 3a -b -c -4d + 4e = 3 a -b - c + 4 \ell,
  \label{ni}
\end{equation}
where we defined $\ell = e - d$, for the sake of simplicity.

In the same spirit, we define an operator which counts the weight density,
\begin{equation*}
  \begin{aligned}
    W( T^{\mu,\nu\lambda} ) &= 1 & W( A_\mu ) &= 0 & W(  \nabla_{\mu} ) &= 0 \\
    W( \epsilon_{\mu\nu\lambda\rho} ) &= -1 & & & W( \epsilon^{\mu\nu\lambda\rho} ) &= 1 .
  \end{aligned}
\end{equation*}
Thus, the weight ($w$) of the general expression above is given by
\begin{equation}
  w = a + \ell.
  \label{wd}
\end{equation}

\subsection{\label{sec:im}Inverse metric density}

Now, we illustrate the usefulness of the \emph{dimensional analysis} by building the most general symmetric $\binom{2}{0}$-tensor density, which we call the \emph{inverse metric density}. This particular case fixes $n = 2$ and $w = 1$.

Equation~\eqref{wd} can solved by choosing either $a=1$ or $\ell = 1$, which imply that Eq.~\eqref{ni} is restricted to
\begin{equation*}
  3 - b - c = 2,
\end{equation*}
or
\begin{equation*}
  4 - b - c = 2,
\end{equation*}
respectively. The former, yields the terms
\begin{equation}
  T^{\mu,\nu\lambda} A_\lambda \text{ and } \nabla_\lambda T^{\mu,\nu\lambda},
\end{equation}
no other contraction of indices is allowed due to the symmetry. The latter, yields another possibility,
\begin{equation}
  \epsilon_{\lambda\kappa\rho\sigma} T^{\mu, \lambda\kappa} T^{\nu, \rho\sigma}.
\end{equation}

Finally, one can check that any other choice to solve Eq.~\eqref{wd}---by allowing negative values of $\ell$---make impossible to solve Eq.~\eqref{ni}. Ergo, there is no other term in a symmetric $\binom{2}{0}$-tensor density built up with these fields. Explicitly, this general tensor density is written in Eq.~\eqref{geng}. The summary of this analysis is presented in Table~\ref{tab:imd}.
\begin{table}
  \caption{Possible terms contributing to the inverse density metric. }
  \label{tab:imd}
  \begin{tabular}{|C{.23\linewidth}C{.23\linewidth}C{.23\linewidth}C{.23\linewidth}|}
    \hline
    $a$ & $b$ & $c$ & $\ell$ \\
    \hline
    1 & 1 & 0 & 0 \\
    1 & 0 & 1 & 0 \\
    2 & 0 & 0 &-1 \\
    0 & 2 & 0 & 1 \\
    0 & 1 & 1 & 1 \\
    0 & 0 & 2 & 1 \\
    \hline
  \end{tabular}
\end{table}

\subsection{\label{sec:ld}Lagrangian density}

\begin{table}
  \caption{Possible terms contributing to the Lagrangian density.}
  \label{tab:ld}
  \begin{tabular}{|C{.23\linewidth}C{.23\linewidth}C{.23\linewidth}C{.23\linewidth}|}
    \hline
    $a$ & $b$ & $c$ & $\ell$ \\
    \hline
    1 & 3 & 0 & 0\\
    1 & 2 & 1 & 0\\
    1 & 1 & 2 & 0\\
    1 & 0 & 3 & 0\\
    2 & 2 & 0 & -1\\
    2 & 1 & 1 & -1\\
    2 & 0 & 2 & -1\\
    3 & 1 & 0 & -2\\
    3 & 0 & 1 & -2\\
    4 & 0 & 0 & -3\\
    0 & 4 & 0 & 1\\
    0 & 3 & 1 & 1\\
    0 & 2 & 2 & 1\\
    0 & 1 & 3 & 1\\
    0 & 0 & 4 & 1\\
    \hline
  \end{tabular}
\end{table}

The work of building up the most general scalar density with our ingredients is as simple as before, but the analysis is much longer. Thus, we will show the procedure with lesser details than before.

First, we are interested in a scalar density, which sets $n = 0$ and $w = 1$ in Eqs.~\eqref{ni} and~\eqref{wd}, i.e.,
\begin{align}
  3a -b -c  + 4\ell &= 0, \label{nib}\\
  a + \ell &= 1. \label{wdb}
\end{align}

A possible solution of Eq~\eqref{wdb} is $a=1$ and $\ell = 0$. Such choice allows four possible solutions of Eq.~\eqref{nib}, but only three of them are nonvanishing. These generate the terms of Eq.~\eqref{4dfull} whose coefficients were called $B_3$, $B_4$, $B_5$, $C_1$, $C_2$, $D_6$, $D_7$ and $E_2$.

If one choose $\ell = 1$ for solving Eq.~\eqref{wdb}, all possible solutions for Eq.~\eqref{nib} are either vanishing or yield topological terms.

The case with $a=2$ and $\ell = -1$ gives three possibilities to solve Eq.~\eqref{nib}. These choices give the terms the Lagrangian density whose coefficients are $B_1$, $B_2$, $D_4$, $D_5$, $E_1$, and $F_4$.

For $a = 3$ and $\ell = -2$ there are only two possible solutions of Eq.~\eqref{nib}, and these yield the terms of the action with coefficients $D_1$, $D_2$, $D_3$ and $F_3$. And also, the choice $a = 4$ and $\ell = -3$ solves the equations, and yield the terms in the action with coefficients $F_1$ and $F_2$.

Notice that for values $a \ge 5$, it is not possible to solve both Eqs.~\eqref{nib} and~\eqref{wdb} simultaneously. Therefore, we conclude that Eq.~\eqref{4dfull} is the most general action built up with these fields. A summary of the choices for building the Lagrangian density is presented in Table~\ref{tab:ld}.

\section{\label{Known}Known metric solutions to the parallel Ricci equations}

In this Appendix, we will present a brief compendium of known (Riemannian) metric solutions of Eq.~\eqref{SimpleEOM} in four dimensions. The results presented in this Appendix is not original, and can be found in Refs.~\cite{gray1978einstein,derdzinski1980classification,derdzinski1982compact,Derdzinski:1985,Besse}.

\paragraph{Einstein spaces.---} The simplest example of a manifold with parallel Ricci is a Riemannian Einstein manifols. This example was explained on the main text.

\paragraph{Riemannian products.---} The Riemamian product of manifolds with harmonic curvature.

\paragraph{Conformally flat manifolds.---} These are solutions in four dimensions if their scalar curvature is constant.

\paragraph{Warped products.---} Compact warped products of the form \mbox{$(S^1 \times \bar{M}, \de{t}^2 + f(t) \de{s}^2(\bar{g}) )$,} where $(\bar{M}, \bar{g})$ is an Einstein manifold with positive scalar curvature, $\bar{R} > 0$, and $f$ is a positive function on $S^1$ satisfying the differential equation
\begin{equation}
  \ddot{f} - \frac{1}{3} \bar{R} = c f,
\end{equation}
for a negative constant $c$.

{Twisted} warped products of the form $(\R \times \bar{M} ) / \Z$ have parallel Ricci, if the $\Z$-action on the product metric involves an isometry of $\bar{g}$.

Compact warped products of the form \mbox{$(M_1 \times M_2, f^{2} \cdot (g_1 \times g_2) )$,} where $M_i$ are two two-dimensional manifolds, $M_1$ has constant Gau\ss{}ian curvature, $K_1 < 0$, and $M_2$ is an Einstein manifold with scalar curvature $R_2 = - 2 K_1$. The function $f: M_1 \to \R^+$ is a $C^\infty$ solution of 
\begin{equation*}
  \Lap f = c \, f^3 \text{ for } c > 0.
\end{equation*}

Many examples and explicit conditions of Riemannian metrics with harmonic curvature, and conditions for them to have parallel Ricci, can be found in Ref.~\cite{derdzinski1988riemannian}.

Although the above are examples of manifolds with parallel Ricci within the context of ``Riemannian'' geometries, examples of connnections with parallel Ricci can be found in symplectic geometries, see Ref.~\cite{cahen2000symplectic}.

\section{\label{Schw}Explicit example: Schwarzschild ansatz, and Birkhoff theorem}



As pointed out, all vacuum solutions of Einstein's equations---with and without cosmological constant---satisfy the parallel Ricci condition trivially~\cite{bourguignon1981varietes}. Nonetheless, an important result in General Relativity is the stability of the Schwarzschild metric, known as the Birkhoff theorem~\cite{Jebsen1921,Birkhoff1923,Alexandrow1923,Eisland1925}, which implies that the gravitational collapse of a spherically symmetric astrophysical object cannot emit gravitational radiation. 

In this section we show that the simplest ansatz of a static and spherically symmetric metric, which solves the Eq.~\eqref{SimpleEOM} is necessarily the (Anti-)de-Sitter--Schwarzschild metric. This should be taken as a first step toward a Birkhoff theorem, despite the fact that a complete proof is far from been achieved.

Consider first the simplest metric ansatz, whose line element is
\begin{equation}
  \de{s}^2 = - f(t,r) \de{t}^2 + \frac{ \de{r}^2 }{ f(t,r) } + r^2 \de{\Omega}^2.
  \label{simpleSchw}
\end{equation}
\begin{widetext}
  The nontrivial field equation for the ansatz~\eqref{simpleSchw} are
  \begin{dmath}
    \nabla_{[\mu} R_{\nu]\lambda} = \frac{1}{2 \, r^{2} f^{3}} \Bigg( r^{2} f^{4} \frac{\partial^3}{\partial r^{3}}f + 2 \, r f^{4} \frac{\partial^2}{\partial r^{2}}f + r^{2} f^{2} \frac{\partial^3}{\partial t^{2}\partial r^{}}f - 4 \, r^{2} f \frac{\partial}{\partial t^{}}f \frac{\partial^2}{\partial t^{}\partial r^{}}f - 2 \, f^{4} \frac{\partial}{\partial r^{}}f + 2 \, {\left(3 \, r^{2} \frac{\partial}{\partial r^{}}f - 2 \, r f\right)} \frac{\partial}{\partial t^{}}f^{2} - 2 \, {\left(r^{2} f \frac{\partial}{\partial r^{}}f - r f^{2}\right)} \frac{\partial^2}{\partial t^{2}}f \Bigg) \,
    \delta_{\mu\nu}^{10} \delta_\lambda^0
    +
    \frac{1}{2 \, r^{2} f^{5}} \Bigg( r^{2} f^{4} \frac{\partial^3}{\partial t^{}\partial r^{2}}f + 2 \, f^{4} \frac{\partial}{\partial t^{}}f + 6 \, r^{2} \frac{\partial}{\partial t^{}}f^{3} - 6 \, r^{2} f \frac{\partial}{\partial t^{}}f \frac{\partial^2}{\partial t^{2}}f + r^{2} f^{2} \frac{\partial^3}{\partial t^{3}}f \Bigg) \,
    \delta_{\mu\nu}^{10} \delta_\lambda^1
    +
    r \frac{\partial^2}{\partial t^{}\partial r^{}}f \,
    \Big( \delta_{\mu\nu}^{20} \delta_\lambda^2 + \sin^2(\theta) \delta_{\mu\nu}^{30} \delta_\lambda^3 \Big)
    +
    \frac{1}{2 \, r f^{3}} \Bigg(r^{2} f^{3} \frac{\partial^2}{\partial r^{2}}f - 2 \, f^{4} + 2 \, r^{2} \frac{\partial}{\partial t^{}}f^{2} - r^{2} f \frac{\partial^2}{\partial t^{2}}f + 2 \, f^{3} \Bigg) \,
    \Big( \delta_{\mu\nu}^{21} \delta_\lambda^2 + \sin^2(\theta) \delta_{\mu\nu}^{31} \delta_\lambda^3 \Big)
    = 0
  \end{dmath}
\end{widetext}
Even though four out of the six nontrivial field equations are independent, the system can be solved uniquely. First, the component along the $(\mu\nu\lambda) = (2,0,2)$ yields $f = T(t) + R(r)$. Then, one notices that the equation along $(\mu\nu\lambda) = (1,0,1)$ contains a term depending solely on $t$, which implies that $\dot{T}$ vanishes, i.e., $T$ is a constant. From here, the equation for $(\mu\nu\lambda) = (2,1,2)$ gets simple enough to be solved and yields, \mbox{$R(r) = 1 - T + \frac{\alpha}{r} + \beta r^2$} or equivalently
\begin{equation}
  f(r) = 1 + \frac{\alpha}{r} + \beta r^2,
  \label{simplef}
\end{equation}
with $\alpha$ and $\beta$ integration constants. Finally, it can be checked that albeit the equation along the components $(\mu,\nu,\lambda) = (1,0,0)$ was not used, it is satisfied by Eq.~\eqref{simplef}.

The above shows that the simplest time-dependent generalization of Schwarzschild metric is not a solution for the field equations of the considered affine model. Unless, the spacetime metric is static and exactly that for a Schwarzschild's exterior solution, with or without cosmological constant.

However, if one tries a metric ansatz with two different factors, say
\begin{equation}
  \de{s}^2 = - A(t,r) \de{t}^2 + \frac{ \de{r}^2 }{ B(t,r) } + r^2 \de{\Omega}^2,
  \label{ABmetric}
\end{equation}
the field equations are so complicated, that although hints on the solubility were found---and yield to Eq.~\eqref{simplef} for both $A$ and $B$---, the uniqueness of the solution has not been proven. The field equations for the metric~\eqref{ABmetric} are 
\begin{widetext}
  \begin{dmath}
    \nabla_{[\mu} R_{\nu]\lambda} = \frac{1}{4 \, r^{2} B^{3} A^{2}} \Bigg( 2 \, r^{2} B^{4} \frac{\partial}{\partial r^{}}A^{3} + 2 \, r^{2} B^{4} A^{2} \frac{\partial^3}{\partial r^{3}}A + 2 \, r^{2} B^{2} A^{2} \frac{\partial^3}{\partial t^{2}\partial r^{}}B - 6 \, r^{2} B A^{2} \frac{\partial}{\partial t^{}}B \frac{\partial^2}{\partial t^{}\partial r^{}}B - r^{2} B^{2} A \frac{\partial}{\partial t^{}}B \frac{\partial^2}{\partial t^{}\partial r^{}}A + 6 \, {\left(r^{2} A^{2} \frac{\partial}{\partial r^{}}B - r B A^{2}\right)} \frac{\partial}{\partial t^{}}B^{2} - 2 \, {\left(r^{2} B^{3} A \frac{\partial}{\partial r^{}}B + r B^{4} A\right)} \frac{\partial}{\partial r^{}}A^{2} - 2 \, {\left(r^{2} B A^{2} \frac{\partial}{\partial r^{}}B - 2 \, r B^{2} A^{2}\right)} \frac{\partial^2}{\partial t^{2}}B - {\left(r^{2} B^{2} A \frac{\partial^2}{\partial t^{}\partial r^{}}B - 2 \, r^{2} B^{2} \frac{\partial}{\partial t^{}}B \frac{\partial}{\partial r^{}}A - {\left(r^{2} B A \frac{\partial}{\partial r^{}}B - 2 \, r B^{2} A\right)} \frac{\partial}{\partial t^{}}B\right)} \frac{\partial}{\partial t^{}}A + {\left(r^{2} B^{3} A^{2} \frac{\partial^2}{\partial r^{2}}B + 2 \, r B^{3} A^{2} \frac{\partial}{\partial r^{}}B - 4 \, B^{4} A^{2} + 3 \, r^{2} B A \frac{\partial}{\partial t^{}}B^{2} - 2 \, r^{2} B^{2} A \frac{\partial^2}{\partial t^{2}}B\right)} \frac{\partial}{\partial r^{}}A + {\left(3 \, r^{2} B^{3} A^{2} \frac{\partial}{\partial r^{}}B - 4 \, r^{2} B^{4} A \frac{\partial}{\partial r^{}}A + 4 \, r B^{4} A^{2}\right)} \frac{\partial^2}{\partial r^{2}}A \Bigg) \,
    \delta_{\mu\nu}^{10} \delta_\lambda^0
    +
    \frac{1}{4 \, r^{2} B^{4} A^{3}} \Bigg( 2 \, r^{2} B^{4} A^{2} \frac{\partial^3}{\partial t^{}\partial r^{2}}A + r^{2} B^{3} A^{2} \frac{\partial^2}{\partial t^{}\partial r^{}}B \frac{\partial}{\partial r^{}}A - r^{2} B^{3} A \frac{\partial}{\partial t^{}}B \frac{\partial}{\partial r^{}}A^{2} + 2 \, r^{2} B^{3} A^{2} \frac{\partial}{\partial t^{}}B \frac{\partial^2}{\partial r^{2}}A + 4 \, B^{3} A^{3} \frac{\partial}{\partial t^{}}B + 6 \, r^{2} A^{2} \frac{\partial}{\partial t^{}}B^{3} - 8 \, r^{2} B A^{2} \frac{\partial}{\partial t^{}}B \frac{\partial^2}{\partial t^{2}}B + 2 \, r^{2} B^{2} A^{2} \frac{\partial^3}{\partial t^{3}}B + 2 \, r^{2} B^{2} \frac{\partial}{\partial t^{}}B \frac{\partial}{\partial t^{}}A^{2} - r^{2} B^{2} A \frac{\partial}{\partial t^{}}B \frac{\partial^2}{\partial t^{2}}A - {\left(r^{2} B^{3} A \frac{\partial}{\partial r^{}}B \frac{\partial}{\partial r^{}}A - 2 \, r^{2} B^{4} \frac{\partial}{\partial r^{}}A^{2} + 2 \, r^{2} B^{4} A \frac{\partial^2}{\partial r^{2}}A - 4 \, r^{2} B A \frac{\partial}{\partial t^{}}B^{2} + 3 \, r^{2} B^{2} A \frac{\partial^2}{\partial t^{2}}B\right)} \frac{\partial}{\partial t^{}}A + {\left(r^{2} B^{3} A^{2} \frac{\partial}{\partial r^{}}B - 2 \, r^{2} B^{4} A \frac{\partial}{\partial r^{}}A\right)} \frac{\partial^2}{\partial t^{}\partial r^{}}A \Bigg) \,
    \delta_{\mu\nu}^{10} \delta_\lambda^1
    +
    \frac{1}{2 \, A^{2}} \Bigg( r A^{2} \frac{\partial^2}{\partial t^{}\partial r^{}}B + r B A \frac{\partial^2}{\partial t^{}\partial r^{}}A + r A \frac{\partial}{\partial t^{}}B \frac{\partial}{\partial r^{}}A - r B \frac{\partial}{\partial t^{}}A \frac{\partial}{\partial r^{}}A \Bigg) \,
    \Big( \delta_{\mu\nu}^{20} \delta_\lambda^2 + \sin^2(\theta) \delta_{\mu\nu}^{30} \delta_\lambda^3 \Big)
    +
    \frac{1}{4 \, r B^{2} A^{2}} \Bigg( 2 \, r^{2} B^{2} A^{2} \frac{\partial^2}{\partial r^{2}}B + r^{2} B^{2} A \frac{\partial}{\partial r^{}}B \frac{\partial}{\partial r^{}}A - r^{2} B^{3} \frac{\partial}{\partial r^{}}A^{2} + 3 \, r^{2} A \frac{\partial}{\partial t^{}}B^{2} - 2 \, r^{2} B A \frac{\partial^2}{\partial t^{2}}B + r^{2} B \frac{\partial}{\partial t^{}}B \frac{\partial}{\partial t^{}}A - 4 \, {\left(B^{3} - B^{2}\right)} A^{2} \Bigg) \,
    \Big( \delta_{\mu\nu}^{21} \delta_\lambda^2 + \sin^2(\theta) \delta_{\mu\nu}^{31} \delta_\lambda^3 \Big)
    = 0,
    \label{hugeeq}
  \end{dmath}
\end{widetext}
which are very complex. Nonetheless, the symmetries in some of the components of these equations seem to favour the relation $A = B$ as solution. In addition, we found a very interesting reference in which a proof of the Birkhoff theorem is given for an Einstein--Cartan theory coupled with the gravitational Yang--Mills~\cite{Ramaswamy:1979zz}, which support the hand-waving symmetry argument above.

Hence, further analysis requires a new strategy. Instead of considering a metric, a connection ansatz is proposed based in the structure of a spherically symmetric metric of the form,
\begin{equation}
  \de{s}^2 = g_{ab}(x) \, \de{x}^a\de{x}^b + e^{2\rho(x)} \, g_{AB}(\vph)  \de{\vph}^A\de{\vph}^B,
  \label{sphere-gen}
\end{equation}
where $x^m$ represent the temporal an radial directions, while $\vph^M$ represent angular coordinates. Notice that the four-dimensional indices have been partitioned as \mbox{$\mu = (a,A)$,} then the general form of the connection is
\begin{dmath}
  \Ga^\lambda{}_{\mu\nu} = \delta^\lambda_a \delta^b_\mu \delta^c_\nu \ga^a{}_{bc} + \delta^\lambda_A \delta^B_\mu \delta^C_\nu \ga^A{}_{BC} + \delta^\lambda_a \delta^B_\mu \delta^C_\nu g_{BC} M^a(x) +  \delta^\lambda_A \delta^A_{(\mu} \delta^c_{\nu)} N_{c}(x),
  \label{connSchw}
\end{dmath}
where $\ga^a{}_{bc}$ and $\ga^A{}_{BC}$ are the Levi-Civita connection for the metrics $g_{ab}$ and $g_{AB}$ respectively, and the factors are \mbox{$M^a(x) = - \tfrac{1}{2}  g^{am}\partial_m e^{2\rho(x)}$} and \mbox{$N_c = 2 \partial_c \rho$}.

From the connection ansatz~\eqref{connSchw} the Ricci tensor is
\begin{dmath}
  R_{\mu\nu}
  = \delta_\mu^m \delta_\nu^n \big( R_{mn} - 2 \nabla_{m} N_n - 2 N_m N_n \big)
  + \delta_\mu^M \delta_\nu^N \big( R_{MN} + g_{MN} \nabla_{a} M^a \big),
\end{dmath}
and the equations of motion are
\begin{widetext}
  \begin{dmath}
    \nabla_{[\lambda} R_{\mu]\nu}
    = \delta_\lambda^l \delta_\mu^m \delta_\nu^n  \bigg( \frac{1}{2} g_{n[m} \partial_{l]} \mathcal{R} + 2 \mathcal{R}  g_{n[m} N_{l]}(x) - 2 \nabla_{[l} \Big( N_{m]} N_n \Big) \bigg)
    + \delta_\lambda^{[l} \delta_\mu^{M]} \delta_\nu^N \, g_{MN} \bigg( \nabla_{l} \nabla_{a} M^a + \frac{\mathcal{R}}{2} g_{la} M^a - 2 N_l - 2 \nabla_{a} \Big( M^a N_l \Big) - 2 M^a N_a N_l \bigg).
    \label{OurEq}
  \end{dmath}
\end{widetext}
We have used that $g_{AB}(\varphi)$ is the metric of a two-sphere, and its curvature tensors are well known. In addition, since the metric $g_{ab}(x)$ is two-dimensional too, its curvature tensors are related with the metric by the relation
\begin{dmath}
  R_{ab}{}^c{}_d = \frac{\mathcal{R}}{2} \Big( \delta^c_a g_{bd} - \delta^c_b g_{ad} \Big).
\end{dmath}

As argue  in Refs.~\cite{bergmann1965spherically,hawking1973large},  the solution depends on the nature of the surfaces described by $\rho = \text{const.}$, which in our parametrization translate to the nature of $N_a$. We first consider the condition $g^{ab} N_a N_b > 0 $, which allows to set $\rho$ as a good radial coordinate. Therefore, the general metric ansatz $g_{ab}$ is diagonal, and the radial direction would be $\rho$, i.e.,
\begin{equation}
  g_{ab} = - \alpha^2(t,\rho) \de{t}_a \otimes \de{t}_b + \beta^2(t,\rho) \de{\rho}_a \otimes \de{\rho}_b.
  \label{metr2d}
\end{equation}
With this ansatz, the nontrivial components of Eq.~\eqref{OurEq} are
\begin{align}
  \beta^2 \partial_{t} \mathcal{R} + 8 \gamma^\rho{}_{t \rho} &= 0,
  \label{eq1}
  \\
  \partial_{\rho} \mathcal{R} - 4 \mathcal{R} + 16 \alpha^{-2} \gamma^\rho{}_{tt} &= 0,
  \label{eq2}
  \\
  \partial_{t} \big( \Delta_x e^{2\rho} \big) + 4 \beta^{-2} e^{2\rho} \gamma^\rho{}_{t \rho} &= 0,
  \label{eq3}
\end{align}
and
\begin{dmath}
  \partial_{\rho} \big( \Delta_x e^{2\rho} \big) + 8 - 4 \Delta_x e^{2\rho} - 16 \beta^{-2} e^{2\rho} - 4 \beta^{-2} e^{2\rho} \gamma^{\rho}{}_{\rho\rho}  + \mathcal{R} e^{2\rho} = 0.
  \label{eq4}
\end{dmath}
From equation~\eqref{eq1} it follows that,
\begin{equation}
  \mathcal{R} - 8 \beta^{-2} = G_1(\rho).
  \label{eq1b}
\end{equation}
In addition, since
\begin{equation}
  \Delta_x e^{2\rho}
  = e^{2\rho} \bigg[ 4 \beta^{-2} + \frac{2}{\alpha \beta} \partial_{\rho} \bigg( \frac{\alpha}{\beta} \bigg) \bigg]
  \equiv  e^{2\rho} \Big[ 4 \beta^{-2} + \mathcal{F} \Big],
\end{equation}
it follows from Eqs.~\eqref{eq3} that,
\begin{equation}
  \mathcal{F} + 2 \beta^{-2} = G_2(\rho).
  \label{eq3b}
\end{equation}

Equations~\eqref{eq1b} and~\eqref{eq3b} implies that all temporal dependence on time within $\mathcal{R}$, $\mathcal{F}$ and $\beta^{-2}$, enters through an additive common function, say $B(t,\rho)$. However, the consistency of these equations with the value of $\mathcal{R}$ calculated from the metric ansatz, in Eq.~\eqref{metr2d}, sets this additive function to zero, $B(t,\rho)$. Therefore, neither $\mathcal{R}$, $\mathcal{F}$ nor $\beta^{-2}$ depend on $t$. By extension---from the definition of $\mathcal{F}$---, $\alpha$ could depend on time through a multiplicative function. Nonetheless, this multiplicative function can be eliminated of the problem by a redefinition of the \emph{time} coordinate.

At this stage, we have proved that \emph{any spherically symmetric (metric) solution of the Eqs.~\eqref{SimpleEOM} is necessarily static}, which is one of the main statements of Birkhoff's theorem.

This result, simplifies the Eq.~\eqref{hugeeq} but the equations are still non-linear. However, one can get ride off the non-linearity by setting $A(r) = B(r)$. In this \emph{linearized} regime, according to Eq.~\eqref{simplef} the general (metric) solution is the (Anti-)de-Sitter--Schwarzschild metric.

Although we restricted ourselves to the \emph{linearized} regime of the field equations, it is possible to go a bit further in the nonlinear regime. As a preliminary result~\cite{OCF-future2}, we have found that using the Frobenius methods with a Laurent series expansion, the leading order of the solutions is three-fold in the radial coordinate. These solutions correspond to the pairs given by $(\alpha,\beta)$---$\alpha$ (resp. $\beta)$ is the leading order of the series expansion for the function $A(r)$ (resp. $B(r)$)---equals to $(-1,-1)$, $(2,-2)$ and $(0,-1)$.


\begin{thebibliography}{86}%
\makeatletter
\providecommand \@ifxundefined [1]{%
 \@ifx{#1\undefined}
}%
\providecommand \@ifnum [1]{%
 \ifnum #1\expandafter \@firstoftwo
 \else \expandafter \@secondoftwo
 \fi
}%
\providecommand \@ifx [1]{%
 \ifx #1\expandafter \@firstoftwo
 \else \expandafter \@secondoftwo
 \fi
}%
\providecommand \natexlab [1]{#1}%
\providecommand \enquote  [1]{``#1''}%
\providecommand \bibnamefont  [1]{#1}%
\providecommand \bibfnamefont [1]{#1}%
\providecommand \citenamefont [1]{#1}%
\providecommand \href@noop [0]{\@secondoftwo}%
\providecommand \href [0]{\begingroup \@sanitize@url \@href}%
\providecommand \@href[1]{\@@startlink{#1}\@@href}%
\providecommand \@@href[1]{\endgroup#1\@@endlink}%
\providecommand \@sanitize@url [0]{\catcode `\\12\catcode `\$12\catcode
  `\&12\catcode `\#12\catcode `\^12\catcode `\_12\catcode `\%12\relax}%
\providecommand \@@startlink[1]{}%
\providecommand \@@endlink[0]{}%
\providecommand \url  [0]{\begingroup\@sanitize@url \@url }%
\providecommand \@url [1]{\endgroup\@href {#1}{\urlprefix }}%
\providecommand \urlprefix  [0]{URL }%
\providecommand \Eprint [0]{\href }%
\providecommand \doibase [0]{http://dx.doi.org/}%
\providecommand \selectlanguage [0]{\@gobble}%
\providecommand \bibinfo  [0]{\@secondoftwo}%
\providecommand \bibfield  [0]{\@secondoftwo}%
\providecommand \translation [1]{[#1]}%
\providecommand \BibitemOpen [0]{}%
\providecommand \bibitemStop [0]{}%
\providecommand \bibitemNoStop [0]{.\EOS\space}%
\providecommand \EOS [0]{\spacefactor3000\relax}%
\providecommand \BibitemShut  [1]{\csname bibitem#1\endcsname}%
\let\auto@bib@innerbib\@empty
\bibitem [{\citenamefont {Einstein}(1915)}]{einstein1915feldgleichungen}%
  \BibitemOpen
  \bibfield  {author} {\bibinfo {author} {\bibfnamefont {Albert}\ \bibnamefont
  {Einstein}},\ }\bibfield  {title} {\enquote {\bibinfo {title} {Die
  feldgleichungen der gravitation},}\ }\href@noop {} {\bibfield  {journal}
  {\bibinfo  {journal} {Sitzungsber. preuss. Akad. Wiss.}\ }\textbf {\bibinfo
  {volume} {1}},\ \bibinfo {pages} {844--847} (\bibinfo {year}
  {1915})}\BibitemShut {NoStop}%
\bibitem [{\citenamefont {Will}(2014)}]{Will:2014kxa}%
  \BibitemOpen
  \bibfield  {author} {\bibinfo {author} {\bibfnamefont {Clifford~M.}\
  \bibnamefont {Will}},\ }\bibfield  {title} {\enquote {\bibinfo {title} {{The
  confrontation between Gerenal Relativity and experiment}},}\ }\href {\doibase
  10.12942/lrr-2014-4} {\bibfield  {journal} {\bibinfo  {journal} {Living Rev.
  Rel.}\ }\textbf {\bibinfo {volume} {17}},\ \bibinfo {pages} {4} (\bibinfo
  {year} {2014})},\ \Eprint {http://arxiv.org/abs/1403.7377} {arXiv:1403.7377
  [gr-qc]} \BibitemShut {NoStop}%
\bibitem [{\citenamefont {Abbott}\ \emph {et~al.}(2016)\citenamefont {Abbott}
  \emph {et~al.}}]{Abbott:2016blz}%
  \BibitemOpen
  \bibfield  {author} {\bibinfo {author} {\bibfnamefont {B.~P.}\ \bibnamefont
  {Abbott}} \emph {et~al.} (\bibinfo {collaboration} {{Virgo and LIGO
  scientific}}),\ }\bibfield  {title} {\enquote {\bibinfo {title} {{Observation
  of gravitational waves from a binary black hole merger}},}\ }\href {\doibase
  10.1103/PhysRevLett.116.061102} {\bibfield  {journal} {\bibinfo  {journal}
  {Phys. Rev. Lett.}\ }\textbf {\bibinfo {volume} {116}},\ \bibinfo {pages}
  {061102} (\bibinfo {year} {2016})},\ \Eprint
  {http://arxiv.org/abs/1602.03837} {arXiv:1602.03837 [gr-qc]} \BibitemShut
  {NoStop}%
\bibitem [{\citenamefont {Cartan}(1922)}]{Cartan1922}%
  \BibitemOpen
  \bibfield  {author} {\bibinfo {author} {\bibfnamefont {Elie}\ \bibnamefont
  {Cartan}},\ }\bibfield  {title} {\enquote {\bibinfo {title} {Sur une
  g\'en\'eralisation de la notion de courbure de riemann et les espaces \`a
  torsion},}\ }\href {http://gallica.bnf.fr/ark:/12148/bpt6k3127j.image.langFR}
  {\bibfield  {journal} {\bibinfo  {journal} {C. R. Acad. Sci. Paris}\ }\textbf
  {\bibinfo {volume} {174}},\ \bibinfo {pages} {593} (\bibinfo {year}
  {1922})}\BibitemShut {NoStop}%
\bibitem [{\citenamefont {Cartan}(1923)}]{Cartan1923}%
  \BibitemOpen
  \bibfield  {author} {\bibinfo {author} {\bibfnamefont {Elie}\ \bibnamefont
  {Cartan}},\ }\bibfield  {title} {\enquote {\bibinfo {title} {Sur les
  vari{\'e}t{\'e}s {\`a} connexion affine et la th{\'e}orie de la
  relativit{\'e} g{\'e}n{\'e}ralis{\'e}e (premi{\`e}re partie)},}\ }\href
  {http://archive.numdam.org/article/ASENS_1923_3_40__325_0.pdf} {\bibfield
  {journal} {\bibinfo  {journal} {Ann. Ec. Norm. Super.}\ }\textbf {\bibinfo
  {volume} {40}},\ \bibinfo {pages} {325} (\bibinfo {year} {1923})}\BibitemShut
  {NoStop}%
\bibitem [{\citenamefont {Cartan}(1924)}]{Cartan1924}%
  \BibitemOpen
  \bibfield  {author} {\bibinfo {author} {\bibfnamefont {Elie}\ \bibnamefont
  {Cartan}},\ }\bibfield  {title} {\enquote {\bibinfo {title} {Sur les
  vari\'et\'es \`a connexion affine, et la th\'eorie de la relativit\'e
  g\'en\'eralis\'ee (premi\`ere partie) (suite)},}\ }\href
  {http://www.numdam.org/numdam-bin/item?id=ASENS_1924_3_41__1_0} {\bibfield
  {journal} {\bibinfo  {journal} {Ann. Ec. Norm. Super.}\ }\textbf {\bibinfo
  {volume} {41}},\ \bibinfo {pages} {1} (\bibinfo {year} {1924})}\BibitemShut
  {NoStop}%
\bibitem [{\citenamefont {Cartan}(1925)}]{Cartan1925}%
  \BibitemOpen
  \bibfield  {author} {\bibinfo {author} {\bibfnamefont {Elie}\ \bibnamefont
  {Cartan}},\ }\bibfield  {title} {\enquote {\bibinfo {title} {Sur les
  vari\'et\'es \`a connexion affine et la th\'eorie de la relativit\'e
  g\'en\'eralis\'ee, part ii,},}\ }\href
  {http://www.numdam.org/numdam-bin/item?id=ASENS_1925_3_42__17_0} {\bibfield
  {journal} {\bibinfo  {journal} {Ann. Ec. Norm. Super.}\ }\textbf {\bibinfo
  {volume} {42}},\ \bibinfo {pages} {17} (\bibinfo {year} {1925})}\BibitemShut
  {NoStop}%
\bibitem [{\citenamefont {'t~Hooft}(1973)}]{'tHooft:1973us}%
  \BibitemOpen
  \bibfield  {author} {\bibinfo {author} {\bibfnamefont {Gerard}\ \bibnamefont
  {'t~Hooft}},\ }\bibfield  {title} {\enquote {\bibinfo {title} {{An algorithm
  for the poles at dimension four in the dimensional regularization
  procedure}},}\ }\href {\doibase 10.1016/0550-3213(73)90263-0} {\bibfield
  {journal} {\bibinfo  {journal} {Nucl. Phys. B}\ }\textbf {\bibinfo {volume}
  {62}},\ \bibinfo {pages} {444} (\bibinfo {year} {1973})}\BibitemShut
  {NoStop}%
\bibitem [{\citenamefont {'t~Hooft}\ and\ \citenamefont
  {Veltman}(1974)}]{'tHooft:1974bx}%
  \BibitemOpen
  \bibfield  {author} {\bibinfo {author} {\bibfnamefont {Gerard}\ \bibnamefont
  {'t~Hooft}}\ and\ \bibinfo {author} {\bibfnamefont {M.~J.~G.}\ \bibnamefont
  {Veltman}},\ }\bibfield  {title} {\enquote {\bibinfo {title} {{One loop
  divergencies in the theory of gravitation}},}\ }\href@noop {} {\bibfield
  {journal} {\bibinfo  {journal} {Annales Poincare Phys. Theor. A}\ }\textbf
  {\bibinfo {volume} {20}},\ \bibinfo {pages} {69} (\bibinfo {year}
  {1974})}\BibitemShut {NoStop}%
\bibitem [{\citenamefont {Deser}\ and\ \citenamefont {van
  Nieuwenhuizen}(1974{\natexlab{a}})}]{Deser:1974cz}%
  \BibitemOpen
  \bibfield  {author} {\bibinfo {author} {\bibfnamefont {Stanley}\ \bibnamefont
  {Deser}}\ and\ \bibinfo {author} {\bibfnamefont {P.}~\bibnamefont {van
  Nieuwenhuizen}},\ }\bibfield  {title} {\enquote {\bibinfo {title} {{One Loop
  Divergences of Quantized Einstein-Maxwell Fields}},}\ }\href {\doibase
  10.1103/PhysRevD.10.401} {\bibfield  {journal} {\bibinfo  {journal} {Phys.
  Rev. D}\ }\textbf {\bibinfo {volume} {10}},\ \bibinfo {pages} {401} (\bibinfo
  {year} {1974}{\natexlab{a}})}\BibitemShut {NoStop}%
\bibitem [{\citenamefont {Deser}\ and\ \citenamefont {van
  Nieuwenhuizen}(1974{\natexlab{b}})}]{Deser:1974cy}%
  \BibitemOpen
  \bibfield  {author} {\bibinfo {author} {\bibfnamefont {Stanley}\ \bibnamefont
  {Deser}}\ and\ \bibinfo {author} {\bibfnamefont {P.}~\bibnamefont {van
  Nieuwenhuizen}},\ }\bibfield  {title} {\enquote {\bibinfo {title}
  {{Nonrenormalizability of the Quantized Dirac-Einstein System}},}\ }\href
  {\doibase 10.1103/PhysRevD.10.411} {\bibfield  {journal} {\bibinfo  {journal}
  {Phys. Rev. D}\ }\textbf {\bibinfo {volume} {10}},\ \bibinfo {pages} {411}
  (\bibinfo {year} {1974}{\natexlab{b}})}\BibitemShut {NoStop}%
\bibitem [{\citenamefont {Arnowitt}\ \emph {et~al.}(1959)\citenamefont
  {Arnowitt}, \citenamefont {Deser},\ and\ \citenamefont
  {Misner}}]{Arnowitt:1959ah}%
  \BibitemOpen
  \bibfield  {author} {\bibinfo {author} {\bibfnamefont {Richard~L.}\
  \bibnamefont {Arnowitt}}, \bibinfo {author} {\bibfnamefont {Stanley}\
  \bibnamefont {Deser}}, \ and\ \bibinfo {author} {\bibfnamefont {Charles~W.}\
  \bibnamefont {Misner}},\ }\bibfield  {title} {\enquote {\bibinfo {title}
  {{Dynamical Structure and Definition of Energy in General Relativity}},}\
  }\href {\doibase 10.1103/PhysRev.116.1322} {\bibfield  {journal} {\bibinfo
  {journal} {Phys. Rev.}\ }\textbf {\bibinfo {volume} {116}},\ \bibinfo {pages}
  {1322} (\bibinfo {year} {1959})}\BibitemShut {NoStop}%
\bibitem [{\citenamefont {Arnowitt}\ \emph {et~al.}(1960)\citenamefont
  {Arnowitt}, \citenamefont {Deser},\ and\ \citenamefont
  {Misner}}]{Arnowitt:1960es}%
  \BibitemOpen
  \bibfield  {author} {\bibinfo {author} {\bibfnamefont {Richard~L.}\
  \bibnamefont {Arnowitt}}, \bibinfo {author} {\bibfnamefont {Stanley}\
  \bibnamefont {Deser}}, \ and\ \bibinfo {author} {\bibfnamefont {Charles~W.}\
  \bibnamefont {Misner}},\ }\bibfield  {title} {\enquote {\bibinfo {title}
  {{Canonical variables for general relativity}},}\ }\href {\doibase
  10.1103/PhysRev.117.1595} {\bibfield  {journal} {\bibinfo  {journal} {Phys.
  Rev.}\ }\textbf {\bibinfo {volume} {117}},\ \bibinfo {pages} {1595} (\bibinfo
  {year} {1960})}\BibitemShut {NoStop}%
\bibitem [{\citenamefont {Wheeler}(1964)}]{WheelerGeo}%
  \BibitemOpen
  \bibfield  {author} {\bibinfo {author} {\bibfnamefont {John~A.}\ \bibnamefont
  {Wheeler}},\ }\enquote {\bibinfo {title} {{R}elativity, groups and
  topology},}\ \ (\bibinfo  {publisher} {Gordon and Breach},\ \bibinfo {year}
  {1964})\ Chap.\ \bibinfo {chapter} {{G}eometrodynamics and the issue of the
  final state}, p.\ \bibinfo {pages} {317}\BibitemShut {NoStop}%
\bibitem [{\citenamefont {DeWitt}(1967{\natexlab{a}})}]{DeWitt:1967yk}%
  \BibitemOpen
  \bibfield  {author} {\bibinfo {author} {\bibfnamefont {Bryce~S.}\
  \bibnamefont {DeWitt}},\ }\bibfield  {title} {\enquote {\bibinfo {title}
  {{Quantum Theory of Gravity. 1. The Canonical Theory}},}\ }\href {\doibase
  10.1103/PhysRev.160.1113} {\bibfield  {journal} {\bibinfo  {journal} {Phys.
  Rev.}\ }\textbf {\bibinfo {volume} {160}},\ \bibinfo {pages} {1113} (\bibinfo
  {year} {1967}{\natexlab{a}})}\BibitemShut {NoStop}%
\bibitem [{\citenamefont {DeWitt}(1967{\natexlab{b}})}]{DeWitt:1967ub}%
  \BibitemOpen
  \bibfield  {author} {\bibinfo {author} {\bibfnamefont {Bryce~S.}\
  \bibnamefont {DeWitt}},\ }\bibfield  {title} {\enquote {\bibinfo {title}
  {{Quantum Theory of Gravity. 2. The Manifestly Covariant Theory}},}\ }\href
  {\doibase 10.1103/PhysRev.162.1195} {\bibfield  {journal} {\bibinfo
  {journal} {Phys. Rev.}\ }\textbf {\bibinfo {volume} {162}},\ \bibinfo {pages}
  {1195} (\bibinfo {year} {1967}{\natexlab{b}})}\BibitemShut {NoStop}%
\bibitem [{\citenamefont {DeWitt}(1967{\natexlab{c}})}]{DeWitt:1967uc}%
  \BibitemOpen
  \bibfield  {author} {\bibinfo {author} {\bibfnamefont {Bryce~S.}\
  \bibnamefont {DeWitt}},\ }\bibfield  {title} {\enquote {\bibinfo {title}
  {{Quantum Theory of Gravity. 3. Applications of the Covariant Theory}},}\
  }\href {\doibase 10.1103/PhysRev.162.1239} {\bibfield  {journal} {\bibinfo
  {journal} {Phys. Rev.}\ }\textbf {\bibinfo {volume} {162}},\ \bibinfo {pages}
  {1239} (\bibinfo {year} {1967}{\natexlab{c}})}\BibitemShut {NoStop}%
\bibitem [{\citenamefont {Gambini}\ and\ \citenamefont
  {Trias}(1981)}]{Gambini:1980yz}%
  \BibitemOpen
  \bibfield  {author} {\bibinfo {author} {\bibfnamefont {R.}~\bibnamefont
  {Gambini}}\ and\ \bibinfo {author} {\bibfnamefont {A.}~\bibnamefont
  {Trias}},\ }\bibfield  {title} {\enquote {\bibinfo {title} {{On the
  Geometrical Origin of Gauge Theories}},}\ }\href {\doibase
  10.1103/PhysRevD.23.553} {\bibfield  {journal} {\bibinfo  {journal} {Phys.
  Rev. D}\ }\textbf {\bibinfo {volume} {23}},\ \bibinfo {pages} {553} (\bibinfo
  {year} {1981})}\BibitemShut {NoStop}%
\bibitem [{\citenamefont {Gambini}\ and\ \citenamefont
  {Trias}(1986)}]{Gambini:1986ew}%
  \BibitemOpen
  \bibfield  {author} {\bibinfo {author} {\bibfnamefont {Rodolfo}\ \bibnamefont
  {Gambini}}\ and\ \bibinfo {author} {\bibfnamefont {Antoni}\ \bibnamefont
  {Trias}},\ }\bibfield  {title} {\enquote {\bibinfo {title} {{Gauge Dynamics
  in the C Representation}},}\ }\href {\doibase 10.1016/0550-3213(86)90221-X}
  {\bibfield  {journal} {\bibinfo  {journal} {Nucl. Phys. B}\ }\textbf
  {\bibinfo {volume} {278}},\ \bibinfo {pages} {436} (\bibinfo {year}
  {1986})}\BibitemShut {NoStop}%
\bibitem [{\citenamefont {Ashtekar}(1986)}]{Ashtekar:1986yd}%
  \BibitemOpen
  \bibfield  {author} {\bibinfo {author} {\bibfnamefont {A.}~\bibnamefont
  {Ashtekar}},\ }\bibfield  {title} {\enquote {\bibinfo {title} {{New Variables
  for Classical and Quantum Gravity}},}\ }\href {\doibase
  10.1103/PhysRevLett.57.2244} {\bibfield  {journal} {\bibinfo  {journal}
  {Phys. Rev. Lett.}\ }\textbf {\bibinfo {volume} {57}},\ \bibinfo {pages}
  {2244--2247} (\bibinfo {year} {1986})}\BibitemShut {NoStop}%
\bibitem [{\citenamefont {Ashtekar}(1987)}]{Ashtekar:1987gu}%
  \BibitemOpen
  \bibfield  {author} {\bibinfo {author} {\bibfnamefont {A.}~\bibnamefont
  {Ashtekar}},\ }\bibfield  {title} {\enquote {\bibinfo {title} {{New
  Hamiltonian Formulation of General Relativity}},}\ }\href {\doibase
  10.1103/PhysRevD.36.1587} {\bibfield  {journal} {\bibinfo  {journal} {Phys.
  Rev. D}\ }\textbf {\bibinfo {volume} {36}},\ \bibinfo {pages} {1587}
  (\bibinfo {year} {1987})}\BibitemShut {NoStop}%
\bibitem [{\citenamefont {Holst}(1996)}]{Holst:1995pc}%
  \BibitemOpen
  \bibfield  {author} {\bibinfo {author} {\bibfnamefont {Soren}\ \bibnamefont
  {Holst}},\ }\bibfield  {title} {\enquote {\bibinfo {title} {{Barbero's
  Hamiltonian derived from a generalized Hilbert-Palatini action}},}\ }\href
  {\doibase 10.1103/PhysRevD.53.5966} {\bibfield  {journal} {\bibinfo
  {journal} {Phys. Rev. D}\ }\textbf {\bibinfo {volume} {53}},\ \bibinfo
  {pages} {5966} (\bibinfo {year} {1996})},\ \Eprint
  {http://arxiv.org/abs/gr-qc/9511026} {arXiv:gr-qc/9511026 [gr-qc]}
  \BibitemShut {NoStop}%
\bibitem [{\citenamefont {Kibble}(1961)}]{Kibble:1961ba}%
  \BibitemOpen
  \bibfield  {author} {\bibinfo {author} {\bibfnamefont {T.~W.~B.}\
  \bibnamefont {Kibble}},\ }\bibfield  {title} {\enquote {\bibinfo {title}
  {{Lorentz invariance and the gravitational field}},}\ }\href {\doibase
  10.1063/1.1703702} {\bibfield  {journal} {\bibinfo  {journal} {J. Math.
  Phys.}\ }\textbf {\bibinfo {volume} {2}},\ \bibinfo {pages} {212--221}
  (\bibinfo {year} {1961})}\BibitemShut {NoStop}%
\bibitem [{\citenamefont {Hehl}\ \emph {et~al.}(1976)\citenamefont {Hehl},
  \citenamefont {von~der Heyde}, \citenamefont {Kerlick},\ and\ \citenamefont
  {Nester}}]{Hehl:1976kj}%
  \BibitemOpen
  \bibfield  {author} {\bibinfo {author} {\bibfnamefont {Friedrich~W.}\
  \bibnamefont {Hehl}}, \bibinfo {author} {\bibfnamefont {Paul}\ \bibnamefont
  {von~der Heyde}}, \bibinfo {author} {\bibfnamefont {G.~David}\ \bibnamefont
  {Kerlick}}, \ and\ \bibinfo {author} {\bibfnamefont {James~M.}\ \bibnamefont
  {Nester}},\ }\bibfield  {title} {\enquote {\bibinfo {title} {General
  relativity with spin and torsion: Foundations and prospects},}\ }\href
  {\doibase 10.1103/RevModPhys.48.393} {\bibfield  {journal} {\bibinfo
  {journal} {Rev. Mod. Phys.}\ }\textbf {\bibinfo {volume} {48}},\ \bibinfo
  {pages} {393} (\bibinfo {year} {1976})}\BibitemShut {NoStop}%
\bibitem [{\citenamefont {Shapiro}(2002)}]{Shapiro:2001rz}%
  \BibitemOpen
  \bibfield  {author} {\bibinfo {author} {\bibfnamefont {I.~L.}\ \bibnamefont
  {Shapiro}},\ }\bibfield  {title} {\enquote {\bibinfo {title} {{Physical
  aspects of the space-time torsion}},}\ }\href {\doibase
  10.1016/S0370-1573(01)00030-8} {\bibfield  {journal} {\bibinfo  {journal}
  {Phys. Rep.}\ }\textbf {\bibinfo {volume} {357}},\ \bibinfo {pages} {113}
  (\bibinfo {year} {2002})},\ \Eprint {http://arxiv.org/abs/hep-th/0103093}
  {arXiv:hep-th/0103093 [hep-th]} \BibitemShut {NoStop}%
\bibitem [{\citenamefont {Hammond}(2002)}]{Hammond:2002rm}%
  \BibitemOpen
  \bibfield  {author} {\bibinfo {author} {\bibfnamefont {R.~T.}\ \bibnamefont
  {Hammond}},\ }\bibfield  {title} {\enquote {\bibinfo {title} {{Torsion
  gravity}},}\ }\href {\doibase 10.1088/0034-4885/65/5/201} {\bibfield
  {journal} {\bibinfo  {journal} {Rept. Prog. Phys.}\ }\textbf {\bibinfo
  {volume} {65}},\ \bibinfo {pages} {599} (\bibinfo {year} {2002})}\BibitemShut
  {NoStop}%
\bibitem [{\citenamefont {Hehl}\ \emph {et~al.}(1995)\citenamefont {Hehl},
  \citenamefont {McCrea}, \citenamefont {Mielke},\ and\ \citenamefont
  {Ne'eman}}]{Hehl:1994ue}%
  \BibitemOpen
  \bibfield  {author} {\bibinfo {author} {\bibfnamefont {Friedrich~W.}\
  \bibnamefont {Hehl}}, \bibinfo {author} {\bibfnamefont {J.~Dermott}\
  \bibnamefont {McCrea}}, \bibinfo {author} {\bibfnamefont {Eckehard~W.}\
  \bibnamefont {Mielke}}, \ and\ \bibinfo {author} {\bibfnamefont {Yuval}\
  \bibnamefont {Ne'eman}},\ }\bibfield  {title} {\enquote {\bibinfo {title}
  {{Metric affine gauge theory of gravity: Field equations, Noether identities,
  world spinors, and breaking of dilation invariance}},}\ }\href {\doibase
  10.1016/0370-1573(94)00111-F} {\bibfield  {journal} {\bibinfo  {journal}
  {Phys. Rep.}\ }\textbf {\bibinfo {volume} {258}},\ \bibinfo {pages} {1--171}
  (\bibinfo {year} {1995})},\ \Eprint {http://arxiv.org/abs/gr-qc/9402012}
  {arXiv:gr-qc/9402012 [gr-qc]} \BibitemShut {NoStop}%
\bibitem [{\citenamefont {Pagani}\ and\ \citenamefont
  {Percacci}(2015)}]{Pagani:2015ema}%
  \BibitemOpen
  \bibfield  {author} {\bibinfo {author} {\bibfnamefont {Carlo}\ \bibnamefont
  {Pagani}}\ and\ \bibinfo {author} {\bibfnamefont {Roberto}\ \bibnamefont
  {Percacci}},\ }\bibfield  {title} {\enquote {\bibinfo {title} {{Quantum
  gravity with torsion and non-metricity}},}\ }\href {\doibase
  10.1088/0264-9381/32/19/195019} {\bibfield  {journal} {\bibinfo  {journal}
  {Class. Quant. Grav.}\ }\textbf {\bibinfo {volume} {32}},\ \bibinfo {pages}
  {195019} (\bibinfo {year} {2015})},\ \Eprint
  {http://arxiv.org/abs/1506.02882} {arXiv:1506.02882 [gr-qc]} \BibitemShut
  {NoStop}%
\bibitem [{\citenamefont {Sezgin}\ and\ \citenamefont {van
  Nieuwenhuizen}(1980)}]{Sezgin:1979zf}%
  \BibitemOpen
  \bibfield  {author} {\bibinfo {author} {\bibfnamefont {E.}~\bibnamefont
  {Sezgin}}\ and\ \bibinfo {author} {\bibfnamefont {P.}~\bibnamefont {van
  Nieuwenhuizen}},\ }\bibfield  {title} {\enquote {\bibinfo {title} {{New Ghost
  Free Gravity Lagrangians with Propagating Torsion}},}\ }\href {\doibase
  10.1103/PhysRevD.21.3269} {\bibfield  {journal} {\bibinfo  {journal} {Phys.
  Rev. D}\ }\textbf {\bibinfo {volume} {21}},\ \bibinfo {pages} {3269}
  (\bibinfo {year} {1980})}\BibitemShut {NoStop}%
\bibitem [{\citenamefont {Mannheim}\ and\ \citenamefont
  {Davidson}(2005)}]{Mannheim:2004qz}%
  \BibitemOpen
  \bibfield  {author} {\bibinfo {author} {\bibfnamefont {Philip~D.}\
  \bibnamefont {Mannheim}}\ and\ \bibinfo {author} {\bibfnamefont {Aharon}\
  \bibnamefont {Davidson}},\ }\bibfield  {title} {\enquote {\bibinfo {title}
  {{Dirac quantization of the Pais-Uhlenbeck fourth order oscillator}},}\
  }\href {\doibase 10.1103/PhysRevA.71.042110} {\bibfield  {journal} {\bibinfo
  {journal} {Phys. Rev. A}\ }\textbf {\bibinfo {volume} {71}},\ \bibinfo
  {pages} {042110} (\bibinfo {year} {2005})},\ \Eprint
  {http://arxiv.org/abs/hep-th/0408104} {arXiv:hep-th/0408104 [hep-th]}
  \BibitemShut {NoStop}%
\bibitem [{\citenamefont {Bender}\ and\ \citenamefont
  {Mannheim}(2008{\natexlab{a}})}]{Bender:2007wu}%
  \BibitemOpen
  \bibfield  {author} {\bibinfo {author} {\bibfnamefont {Carl~M.}\ \bibnamefont
  {Bender}}\ and\ \bibinfo {author} {\bibfnamefont {Philip~D.}\ \bibnamefont
  {Mannheim}},\ }\bibfield  {title} {\enquote {\bibinfo {title} {{No-ghost
  theorem for the fourth-order derivative Pais-Uhlenbeck oscillator model}},}\
  }\href {\doibase 10.1103/PhysRevLett.100.110402} {\bibfield  {journal}
  {\bibinfo  {journal} {Phys. Rev. Lett.}\ }\textbf {\bibinfo {volume} {100}},\
  \bibinfo {pages} {110402} (\bibinfo {year} {2008}{\natexlab{a}})},\ \Eprint
  {http://arxiv.org/abs/0706.0207} {arXiv:0706.0207 [hep-th]} \BibitemShut
  {NoStop}%
\bibitem [{\citenamefont {Smilga}(2009)}]{Smilga:2008pr}%
  \BibitemOpen
  \bibfield  {author} {\bibinfo {author} {\bibfnamefont {A.~V.}\ \bibnamefont
  {Smilga}},\ }\bibfield  {title} {\enquote {\bibinfo {title} {{Comments on the
  dynamics of the Pais-Uhlenbeck oscillator}},}\ }\bibfield  {booktitle} {\emph
  {\bibinfo {booktitle} {{Proceedings, 7th Workshop on Quantum Physics with
  Non-Hermitian Operators (PHHQP VII)}}},\ }\href {\doibase
  10.3842/Sigma.2009.017} {\bibfield  {journal} {\bibinfo  {journal} {SIGMA}\
  }\textbf {\bibinfo {volume} {5}},\ \bibinfo {pages} {017} (\bibinfo {year}
  {2009})},\ \Eprint {http://arxiv.org/abs/0808.0139} {arXiv:0808.0139
  [quant-ph]} \BibitemShut {NoStop}%
\bibitem [{\citenamefont {Ilhan}\ and\ \citenamefont
  {Kovner}(2013)}]{Ilhan:2013xe}%
  \BibitemOpen
  \bibfield  {author} {\bibinfo {author} {\bibfnamefont {Ibrahim~Burak}\
  \bibnamefont {Ilhan}}\ and\ \bibinfo {author} {\bibfnamefont {Alex}\
  \bibnamefont {Kovner}},\ }\bibfield  {title} {\enquote {\bibinfo {title}
  {{Some Comments on Ghosts and Unitarity: The Pais-Uhlenbeck Oscillator
  Revisited}},}\ }\href {\doibase 10.1103/PhysRevD.88.044045} {\bibfield
  {journal} {\bibinfo  {journal} {Phys. Rev. D}\ }\textbf {\bibinfo {volume}
  {88}},\ \bibinfo {pages} {044045} (\bibinfo {year} {2013})},\ \Eprint
  {http://arxiv.org/abs/1301.4879} {arXiv:1301.4879 [hep-th]} \BibitemShut
  {NoStop}%
\bibitem [{\citenamefont {Robert}\ and\ \citenamefont
  {Smilga}(2008)}]{Robert:2008}%
  \BibitemOpen
  \bibfield  {author} {\bibinfo {author} {\bibfnamefont {D.}~\bibnamefont
  {Robert}}\ and\ \bibinfo {author} {\bibfnamefont {A.~V.}\ \bibnamefont
  {Smilga}},\ }\bibfield  {title} {\enquote {\bibinfo {title} {{Supersymmetry
  versus ghosts}},}\ }\href {\doibase 10.1063/1.2904474} {\bibfield  {journal}
  {\bibinfo  {journal} {J. Math. Phys.}\ }\textbf {\bibinfo {volume} {49}},\
  \bibinfo {eid} {042104} (\bibinfo {year} {2008}),\
  10.1063/1.2904474}\BibitemShut {NoStop}%
\bibitem [{\citenamefont {Castillo-Felisola}\ and\ \citenamefont
  {Skirzewski}(2015)}]{Skirzewski:2014eta}%
  \BibitemOpen
  \bibfield  {author} {\bibinfo {author} {\bibfnamefont {Oscar}\ \bibnamefont
  {Castillo-Felisola}}\ and\ \bibinfo {author} {\bibfnamefont {Aureliano}\
  \bibnamefont {Skirzewski}},\ }\bibfield  {title} {\enquote {\bibinfo {title}
  {{A polynomial model of purely affine Gravity}},}\ }\href@noop {} {\bibfield
  {journal} {\bibinfo  {journal} {Rev. Mex. Fis.}\ }\textbf {\bibinfo {volume}
  {61}},\ \bibinfo {pages} {421} (\bibinfo {year} {2015})},\ \Eprint
  {http://arxiv.org/abs/1410.6183} {arXiv:1410.6183 [gr-qc]} \BibitemShut
  {NoStop}%
\bibitem [{\citenamefont {Eddington}(1923)}]{Eddington1923math}%
  \BibitemOpen
  \bibfield  {author} {\bibinfo {author} {\bibfnamefont {Arthur~S.}\
  \bibnamefont {Eddington}},\ }\href@noop {} {\emph {\bibinfo {title} {The
  mathematical theory of relativity}}}\ (\bibinfo  {publisher} {Cambridge
  University Press},\ \bibinfo {year} {1923})\BibitemShut {NoStop}%
\bibitem [{\citenamefont {Schr{\"o}dinger}(1950)}]{schrodinger1950space}%
  \BibitemOpen
  \bibfield  {author} {\bibinfo {author} {\bibfnamefont {Erwin}\ \bibnamefont
  {Schr{\"o}dinger}},\ }\href@noop {} {\emph {\bibinfo {title} {Space-time
  structure}}}\ (\bibinfo  {publisher} {Cambridge University Press},\ \bibinfo
  {year} {1950})\BibitemShut {NoStop}%
\bibitem [{\citenamefont {Lewandowski}\ \emph {et~al.}(2006)\citenamefont
  {Lewandowski}, \citenamefont {Okolow}, \citenamefont {Sahlmann},\ and\
  \citenamefont {Thiemann}}]{Lewandowski:2005jk}%
  \BibitemOpen
  \bibfield  {author} {\bibinfo {author} {\bibfnamefont {Jerzy}\ \bibnamefont
  {Lewandowski}}, \bibinfo {author} {\bibfnamefont {Andrzej}\ \bibnamefont
  {Okolow}}, \bibinfo {author} {\bibfnamefont {Hanno}\ \bibnamefont
  {Sahlmann}}, \ and\ \bibinfo {author} {\bibfnamefont {Thomas}\ \bibnamefont
  {Thiemann}},\ }\bibfield  {title} {\enquote {\bibinfo {title} {{Uniqueness of
  diffeomorphism invariant states on holonomy-flux algebras}},}\ }\href
  {\doibase 10.1007/s00220-006-0100-7} {\bibfield  {journal} {\bibinfo
  {journal} {Commun. Math. Phys.}\ }\textbf {\bibinfo {volume} {267}},\
  \bibinfo {pages} {703} (\bibinfo {year} {2006})},\ \Eprint
  {http://arxiv.org/abs/gr-qc/0504147} {arXiv:gr-qc/0504147 [gr-qc]}
  \BibitemShut {NoStop}%
\bibitem [{\citenamefont {McGady}\ and\ \citenamefont
  {Rodina}(2014)}]{McGady:2013sga}%
  \BibitemOpen
  \bibfield  {author} {\bibinfo {author} {\bibfnamefont {David~A.}\
  \bibnamefont {McGady}}\ and\ \bibinfo {author} {\bibfnamefont {Laurentiu}\
  \bibnamefont {Rodina}},\ }\bibfield  {title} {\enquote {\bibinfo {title}
  {{Higher-spin massless $S$-matrices in four-dimensions}},}\ }\href {\doibase
  10.1103/PhysRevD.90.084048} {\bibfield  {journal} {\bibinfo  {journal} {Phys.
  Rev. D}\ }\textbf {\bibinfo {volume} {90}},\ \bibinfo {pages} {084048}
  (\bibinfo {year} {2014})},\ \Eprint {http://arxiv.org/abs/1311.2938}
  {arXiv:1311.2938 [hep-th]} \BibitemShut {NoStop}%
\bibitem [{\citenamefont {Camanho}\ \emph {et~al.}(2014)\citenamefont
  {Camanho}, \citenamefont {Edelstein}, \citenamefont {Maldacena},\ and\
  \citenamefont {Zhiboedov}}]{Camanho:2014apa}%
  \BibitemOpen
  \bibfield  {author} {\bibinfo {author} {\bibfnamefont {Xian~O.}\ \bibnamefont
  {Camanho}}, \bibinfo {author} {\bibfnamefont {Jose~D.}\ \bibnamefont
  {Edelstein}}, \bibinfo {author} {\bibfnamefont {Juan}\ \bibnamefont
  {Maldacena}}, \ and\ \bibinfo {author} {\bibfnamefont {Alexander}\
  \bibnamefont {Zhiboedov}},\ }\bibfield  {title} {\enquote {\bibinfo {title}
  {{Causality Constraints on Corrections to the Graviton Three-Point
  Coupling}},}\ }\href@noop {} {\  (\bibinfo {year} {2014})},\ \Eprint
  {http://arxiv.org/abs/1407.5597} {arXiv:1407.5597 [hep-th]} \BibitemShut
  {NoStop}%
\bibitem [{\citenamefont {Diaz}\ \emph {et~al.}(2014)\citenamefont {Diaz},
  \citenamefont {Higuita},\ and\ \citenamefont {Montesinos}}]{Diaz:2014yua}%
  \BibitemOpen
  \bibfield  {author} {\bibinfo {author} {\bibfnamefont {Bogar}\ \bibnamefont
  {Diaz}}, \bibinfo {author} {\bibfnamefont {Daniel}\ \bibnamefont {Higuita}},
  \ and\ \bibinfo {author} {\bibfnamefont {Merced}\ \bibnamefont
  {Montesinos}},\ }\bibfield  {title} {\enquote {\bibinfo {title} {{Lagrangian
  approach to the physical degree of freedom count}},}\ }\href {\doibase
  10.1063/1.4903183} {\bibfield  {journal} {\bibinfo  {journal} {J. Math.
  Phys.}\ }\textbf {\bibinfo {volume} {55}},\ \bibinfo {pages} {122901}
  (\bibinfo {year} {2014})},\ \Eprint {http://arxiv.org/abs/1406.1156}
  {arXiv:1406.1156 [hep-th]} \BibitemShut {NoStop}%
\bibitem [{\citenamefont {Curtright}(1985)}]{Curtright:1980yk}%
  \BibitemOpen
  \bibfield  {author} {\bibinfo {author} {\bibfnamefont {Thomas}\ \bibnamefont
  {Curtright}},\ }\bibfield  {title} {\enquote {\bibinfo {title} {{Generalized
  Gauge Fields}},}\ }\href {\doibase 10.1016/0370-2693(85)91235-3} {\bibfield
  {journal} {\bibinfo  {journal} {Phys. Lett. B}\ }\textbf {\bibinfo {volume}
  {165}},\ \bibinfo {pages} {304} (\bibinfo {year} {1985})}\BibitemShut
  {NoStop}%
\bibitem [{\citenamefont {Buchholz}\ and\ \citenamefont
  {Fredenhagen}(1977)}]{Buchholz:1976hz}%
  \BibitemOpen
  \bibfield  {author} {\bibinfo {author} {\bibfnamefont {D.}~\bibnamefont
  {Buchholz}}\ and\ \bibinfo {author} {\bibfnamefont {K.}~\bibnamefont
  {Fredenhagen}},\ }\bibfield  {title} {\enquote {\bibinfo {title} {{Dilations
  and Interaction}},}\ }\href {\doibase 10.1063/1.523370} {\bibfield  {journal}
  {\bibinfo  {journal} {J. Math. Phys.}\ }\textbf {\bibinfo {volume} {18}},\
  \bibinfo {pages} {1107} (\bibinfo {year} {1977})}\BibitemShut {NoStop}%
\bibitem [{\citenamefont {Castillo-Felisola}\ and\ \citenamefont
  {Skirzewski}()}]{OCF-future3}%
  \BibitemOpen
  \bibfield  {author} {\bibinfo {author} {\bibfnamefont {Oscar}\ \bibnamefont
  {Castillo-Felisola}}\ and\ \bibinfo {author} {\bibfnamefont {Aureliano}\
  \bibnamefont {Skirzewski}},\ }\href@noop {} {\enquote {\bibinfo {title}
  {{M}etric, torsion, non-metricity and background independence in polynomial
  affine gravity},}\ }\bibinfo {note} {(in preparation)}\BibitemShut {NoStop}%
\bibitem [{Note1()}]{Note1}%
  \BibitemOpen
  \bibinfo {note} {Despite the suggestive name, this is a $\protect \genfrac
  {}(){0pt}{2}{0}$-tensor density which does not necessarily satisfy the metric
  conditions. However, for the simple case of the Einstein--Hilbert action, it
  is in fact the inverse metric density.}\BibitemShut {Stop}%
\bibitem [{\citenamefont {Pop{\l}awski}(2014)}]{Poplawski:2012bw}%
  \BibitemOpen
  \bibfield  {author} {\bibinfo {author} {\bibfnamefont {Nikodem~J.}\
  \bibnamefont {Pop{\l}awski}},\ }\bibfield  {title} {\enquote {\bibinfo
  {title} {{Affine theory of gravitation}},}\ }\href {\doibase
  10.1007/s10714-013-1625-7} {\bibfield  {journal} {\bibinfo  {journal} {Gen.
  Rel. Grav.}\ }\textbf {\bibinfo {volume} {46}},\ \bibinfo {pages} {1625}
  (\bibinfo {year} {2014})},\ \Eprint {http://arxiv.org/abs/1203.0294}
  {arXiv:1203.0294 [gr-qc]} \BibitemShut {NoStop}%
\bibitem [{Note2()}]{Note2}%
  \BibitemOpen
  \bibinfo {note} {We cross-check our result using the software Cadabra~\cite
  {Peeters2007550,peeters2007symbolic,Peeters:2007wn}.}\BibitemShut {Stop}%
\bibitem [{\citenamefont {Nomizu}\ and\ \citenamefont
  {Sasaki}(1994)}]{nomizu1994affine}%
  \BibitemOpen
  \bibfield  {author} {\bibinfo {author} {\bibfnamefont {Katsumi}\ \bibnamefont
  {Nomizu}}\ and\ \bibinfo {author} {\bibfnamefont {Takeshi}\ \bibnamefont
  {Sasaki}},\ }\href@noop {} {\emph {\bibinfo {title} {Affine differential
  geometry}}}\ (\bibinfo  {publisher} {Cambridge University Press},\ \bibinfo
  {year} {1994})\BibitemShut {NoStop}%
\bibitem [{\citenamefont {Bryant}()}]{MO-Bryant02}%
  \BibitemOpen
  \bibfield  {author} {\bibinfo {author} {\bibfnamefont {Robert~L.}\
  \bibnamefont {Bryant}},\ }\href@noop {} {\enquote {\bibinfo {title}
  {{S}ymmetries of non-{R}iemannian curvature tensor (answer)},}\ }\bibinfo
  {howpublished} {{MathOverflow}},\ \bibinfo {note}
  {{URL}:~\url{http://mathoverflow.net/a/212794/25356} (visited on
  2015-08-03)}\BibitemShut {NoStop}%
\bibitem [{\citenamefont {Besse}(2007)}]{Besse}%
  \BibitemOpen
  \bibfield  {author} {\bibinfo {author} {\bibfnamefont {Arthur~L.}\
  \bibnamefont {Besse}},\ }\href@noop {} {\emph {\bibinfo {title} {Einstein
  manifolds}}}\ (\bibinfo  {publisher} {Springer},\ \bibinfo {year}
  {2007})\BibitemShut {NoStop}%
\bibitem [{\citenamefont {Bourguignon}(1981)}]{bourguignon1981varietes}%
  \BibitemOpen
  \bibfield  {author} {\bibinfo {author} {\bibfnamefont {Jean-Pierre}\
  \bibnamefont {Bourguignon}},\ }\bibfield  {title} {\enquote {\bibinfo {title}
  {Les vari{\'e}t{\'e}s de dimension 4 {\`a} signature non nulle dont la
  courbure est harmonique sont d'einstein},}\ }\href@noop {} {\bibfield
  {journal} {\bibinfo  {journal} {Inventiones mathematicae}\ }\textbf {\bibinfo
  {volume} {63}},\ \bibinfo {pages} {263} (\bibinfo {year} {1981})}\BibitemShut
  {NoStop}%
\bibitem [{\citenamefont {Berger}\ and\ \citenamefont
  {Ebin}(1969)}]{Berger:1969}%
  \BibitemOpen
  \bibfield  {author} {\bibinfo {author} {\bibfnamefont {M.}~\bibnamefont
  {Berger}}\ and\ \bibinfo {author} {\bibfnamefont {D.}~\bibnamefont {Ebin}},\
  }\bibfield  {title} {\enquote {\bibinfo {title} {{S}ome decomposition of the
  space of symmetric tensors on riemannian manifold},}\ }\href@noop {}
  {\bibfield  {journal} {\bibinfo  {journal} {J. Diff. Geom.}\ }\textbf
  {\bibinfo {volume} {3}} (\bibinfo {year} {1969})}\BibitemShut {NoStop}%
\bibitem [{Note3()}]{Note3}%
  \BibitemOpen
  \bibinfo {note} {A Codazzi tensor is a symmetric $(0,2)$-type tensor, $T$,
  satisfying the condition $D_X T(Y,Z) = D_Y T(X,Z)$~\cite
  {Derdzinski01071983}.}\BibitemShut {Stop}%
\bibitem [{\citenamefont {Derdzi\'nski}(1985)}]{Derdzinski:1985}%
  \BibitemOpen
  \bibfield  {author} {\bibinfo {author} {\bibfnamefont {Andrzej}\ \bibnamefont
  {Derdzi\'nski}},\ }\bibfield  {title} {\enquote {\bibinfo {title} {Riemannian
  manifolds with harmonic curvature},}\ }in\ \href {\doibase
  10.1007/BFb0075087} {\emph {\bibinfo {booktitle} {Global Differential
  Geometry and Global Analysis 1984}}},\ \bibinfo {series} {Lecture Notes in
  Mathematics}, Vol.\ \bibinfo {volume} {1156},\ \bibinfo {editor} {edited by\
  \bibinfo {editor} {\bibfnamefont {Dirk}\ \bibnamefont {Ferus}}, \bibinfo
  {editor} {\bibfnamefont {Robert~B.}\ \bibnamefont {Gardner}}, \bibinfo
  {editor} {\bibfnamefont {Sigurdur}\ \bibnamefont {Helgason}}, \ and\ \bibinfo
  {editor} {\bibfnamefont {Udo}\ \bibnamefont {Simon}}}\ (\bibinfo  {publisher}
  {Springer Berlin Heidelberg},\ \bibinfo {year} {1985})\ p.~\bibinfo {pages}
  {74}\BibitemShut {NoStop}%
\bibitem [{\citenamefont {Bourguignon}\ and\ \citenamefont
  {Lawson~Jr}(1982)}]{bourguignon1982yang}%
  \BibitemOpen
  \bibfield  {author} {\bibinfo {author} {\bibfnamefont {Jean-Pierre}\
  \bibnamefont {Bourguignon}}\ and\ \bibinfo {author} {\bibfnamefont
  {H.~Blaine}\ \bibnamefont {Lawson~Jr}},\ }\bibfield  {title} {\enquote
  {\bibinfo {title} {Yang-mills theory: its physical origins and differential
  geometric aspects},}\ }in\ \href@noop {} {\emph {\bibinfo {booktitle}
  {Seminar on differential Geometry}}},\ Vol.\ \bibinfo {volume} {102}\
  (\bibinfo {organization} {Annals of Mathematics Studies},\ \bibinfo {year}
  {1982})\ p.\ \bibinfo {pages} {395}\BibitemShut {NoStop}%
\bibitem [{\citenamefont {Nakahara}(2005)}]{Nakahara}%
  \BibitemOpen
  \bibfield  {author} {\bibinfo {author} {\bibfnamefont {Mikio}\ \bibnamefont
  {Nakahara}},\ }\href@noop {} {\emph {\bibinfo {title} {Geometry, Topology and
  Physics}}}\ (\bibinfo  {publisher} {Institute Of Physics},\ \bibinfo {year}
  {2005})\BibitemShut {NoStop}%
\bibitem [{\citenamefont {Stephenson}(1958)}]{stephenson1958quadratic}%
  \BibitemOpen
  \bibfield  {author} {\bibinfo {author} {\bibfnamefont {G.}~\bibnamefont
  {Stephenson}},\ }\bibfield  {title} {\enquote {\bibinfo {title} {Quadratic
  lagrangians and general relativity},}\ }\href {\doibase 10.1007/BF02724929}
  {\bibfield  {journal} {\bibinfo  {journal} {Il Nuovo Cimento Series 10}\
  }\textbf {\bibinfo {volume} {9}},\ \bibinfo {pages} {263--269} (\bibinfo
  {year} {1958})}\BibitemShut {NoStop}%
\bibitem [{\citenamefont {Kilmister}\ and\ \citenamefont
  {Newman}(1961)}]{kilmister1961use}%
  \BibitemOpen
  \bibfield  {author} {\bibinfo {author} {\bibfnamefont {C.~W.}\ \bibnamefont
  {Kilmister}}\ and\ \bibinfo {author} {\bibfnamefont {D.~J.}\ \bibnamefont
  {Newman}},\ }\bibfield  {title} {\enquote {\bibinfo {title} {The use of
  algebraic structures in physics},}\ }in\ \href {\doibase
  10.1017/S0305004100036008} {\emph {\bibinfo {booktitle} {Mathematical
  Proceedings of the Cambridge Philosophical Society}}},\ Vol.~\bibinfo
  {volume} {57}\ (\bibinfo  {publisher} {Cambridge University Press},\ \bibinfo
  {year} {1961})\ p.\ \bibinfo {pages} {851}\BibitemShut {NoStop}%
\bibitem [{\citenamefont {Yang}(1974)}]{Yang1974}%
  \BibitemOpen
  \bibfield  {author} {\bibinfo {author} {\bibfnamefont {C.~N.}\ \bibnamefont
  {Yang}},\ }\bibfield  {title} {\enquote {\bibinfo {title} {Integral formalism
  for gauge fields},}\ }\href {\doibase 10.1103/PhysRevLett.33.445} {\bibfield
  {journal} {\bibinfo  {journal} {Phys. Rev. Lett.}\ }\textbf {\bibinfo
  {volume} {33}},\ \bibinfo {pages} {445} (\bibinfo {year} {1974})}\BibitemShut
  {NoStop}%
\bibitem [{\citenamefont {Zanelli}()}]{JZcomm}%
  \BibitemOpen
  \bibfield  {author} {\bibinfo {author} {\bibfnamefont {Jorge}\ \bibnamefont
  {Zanelli}},\ }\href@noop {} {\enquote {\bibinfo {title} {Private
  communication},}\ }\bibinfo {note} {We thank J.~Zanelli for sharing his
  personal communications with C.~N.~Yang, and clarify this aspect of the
  effective model.}\BibitemShut {Stop}%
\bibitem [{\citenamefont {Chen}(2010)}]{Chen:2010at}%
  \BibitemOpen
  \bibfield  {author} {\bibinfo {author} {\bibfnamefont {Pisin}\ \bibnamefont
  {Chen}},\ }\bibfield  {title} {\enquote {\bibinfo {title} {{Gauge Theory of
  Gravity with de Sitter Symmetry as a Solution to the Cosmological Constant
  Problem and the Dark Energy Puzzle}},}\ }\href {\doibase
  10.1142/9789814335614_0029, 10.1142/S0217732310034274} {\bibfield  {journal}
  {\bibinfo  {journal} {Mod. Phys. Lett. A}\ }\textbf {\bibinfo {volume}
  {25}},\ \bibinfo {pages} {2795} (\bibinfo {year} {2010})},\ \Eprint
  {http://arxiv.org/abs/1002.4275} {arXiv:1002.4275 [gr-qc]} \BibitemShut
  {NoStop}%
\bibitem [{\citenamefont {Kleinert}(1987)}]{Kleinert:1987eb}%
  \BibitemOpen
  \bibfield  {author} {\bibinfo {author} {\bibfnamefont {H.}~\bibnamefont
  {Kleinert}},\ }\bibfield  {title} {\enquote {\bibinfo {title} {{Spontaneous
  quantum gravity: a soluble model}},}\ }\href {\doibase
  10.1016/0370-2693(87)90747-7} {\bibfield  {journal} {\bibinfo  {journal}
  {Phys. Lett. B}\ }\textbf {\bibinfo {volume} {196}},\ \bibinfo {pages} {355}
  (\bibinfo {year} {1987})}\BibitemShut {NoStop}%
\bibitem [{\citenamefont {Bender}\ and\ \citenamefont
  {Mannheim}(2008{\natexlab{b}})}]{Bender:2008vh}%
  \BibitemOpen
  \bibfield  {author} {\bibinfo {author} {\bibfnamefont {Carl~M.}\ \bibnamefont
  {Bender}}\ and\ \bibinfo {author} {\bibfnamefont {Philip~D.}\ \bibnamefont
  {Mannheim}},\ }\bibfield  {title} {\enquote {\bibinfo {title} {{Giving up the
  ghost}},}\ }\href {\doibase 10.1088/1751-8113/41/30/304018} {\bibfield
  {journal} {\bibinfo  {journal} {J. Phys. A}\ }\textbf {\bibinfo {volume}
  {41}},\ \bibinfo {pages} {304018} (\bibinfo {year} {2008}{\natexlab{b}})},\
  \Eprint {http://arxiv.org/abs/0807.2607} {arXiv:0807.2607 [hep-th]}
  \BibitemShut {NoStop}%
\bibitem [{\citenamefont {Mannheim}(2013)}]{Mannheim:2009zj}%
  \BibitemOpen
  \bibfield  {author} {\bibinfo {author} {\bibfnamefont {Philip~D.}\
  \bibnamefont {Mannheim}},\ }\bibfield  {title} {\enquote {\bibinfo {title}
  {{PT symmetry as a necessary and sufficient condition for unitary time
  evolution}},}\ }\href {\doibase 10.1098/rsta.2012.0060} {\bibfield  {journal}
  {\bibinfo  {journal} {Phil. Trans. Roy. Soc. Lond. A}\ }\textbf {\bibinfo
  {volume} {371}},\ \bibinfo {pages} {20120060} (\bibinfo {year} {2013})},\
  \Eprint {http://arxiv.org/abs/0912.2635} {arXiv:0912.2635 [hep-th]}
  \BibitemShut {NoStop}%
\bibitem [{\citenamefont {Bekenstein}\ and\ \citenamefont
  {Majhi}(2015)}]{Bekenstein:2014uwa}%
  \BibitemOpen
  \bibfield  {author} {\bibinfo {author} {\bibfnamefont {Jacob~D.}\
  \bibnamefont {Bekenstein}}\ and\ \bibinfo {author} {\bibfnamefont
  {Bibhas~Ranjan}\ \bibnamefont {Majhi}},\ }\bibfield  {title} {\enquote
  {\bibinfo {title} {{Is the principle of least action a must?}}}\ }\href
  {\doibase 10.1016/j.nuclphysb.2015.01.015} {\bibfield  {journal} {\bibinfo
  {journal} {Nucl. Phys. B}\ }\textbf {\bibinfo {volume} {892}},\ \bibinfo
  {pages} {337} (\bibinfo {year} {2015})},\ \Eprint
  {http://arxiv.org/abs/1411.2424} {arXiv:1411.2424 [hep-th]} \BibitemShut
  {NoStop}%
\bibitem [{\citenamefont {Weinberg}(1972)}]{weinberg1972gravitation}%
  \BibitemOpen
  \bibfield  {author} {\bibinfo {author} {\bibfnamefont {Steven}\ \bibnamefont
  {Weinberg}},\ }\href@noop {} {\emph {\bibinfo {title} {Gravitation and
  cosmology: principles and applications of the general theory of
  relativity}}}\ (\bibinfo  {publisher} {Wiley},\ \bibinfo {year}
  {1972})\BibitemShut {NoStop}%
\bibitem [{\citenamefont {Castillo-Felisola}\ \emph {et~al.}()\citenamefont
  {Castillo-Felisola}, \citenamefont {Orellana},\ and\ \citenamefont
  {Skirzewski}}]{OCF-future2}%
  \BibitemOpen
  \bibfield  {author} {\bibinfo {author} {\bibfnamefont {Oscar}\ \bibnamefont
  {Castillo-Felisola}}, \bibinfo {author} {\bibfnamefont {Oscar}\ \bibnamefont
  {Orellana}}, \ and\ \bibinfo {author} {\bibfnamefont {Aureliano}\
  \bibnamefont {Skirzewski}},\ }\href@noop {} {\enquote {\bibinfo {title}
  {Spherically symmetric solutions to the polynomial affine gravity},}\
  }\bibinfo {note} {(in preparation)}\BibitemShut {NoStop}%
\bibitem [{\citenamefont {Chen}(2014)}]{Chen:2013kia}%
  \BibitemOpen
  \bibfield  {author} {\bibinfo {author} {\bibfnamefont {Pisin}\ \bibnamefont
  {Chen}},\ }\bibfield  {title} {\enquote {\bibinfo {title} {{Recent Progress
  in Cosmology and Particle Astrophysics}},}\ }\href {\doibase
  10.7566/JPSCP.1.011002} {\bibfield  {journal} {\bibinfo  {journal} {JPS Conf.
  Proc.}\ }\textbf {\bibinfo {volume} {1}},\ \bibinfo {pages} {011002}
  (\bibinfo {year} {2014})},\ \Eprint {http://arxiv.org/abs/1310.1107}
  {arXiv:1310.1107 [hep-ph]} \BibitemShut {NoStop}%
\bibitem [{\citenamefont {Chen}\ \emph {et~al.}(2013)\citenamefont {Chen},
  \citenamefont {Izumi},\ and\ \citenamefont {Tung}}]{Chen:2013ota}%
  \BibitemOpen
  \bibfield  {author} {\bibinfo {author} {\bibfnamefont {Pisin}\ \bibnamefont
  {Chen}}, \bibinfo {author} {\bibfnamefont {Keisuke}\ \bibnamefont {Izumi}}, \
  and\ \bibinfo {author} {\bibfnamefont {Nien-En}\ \bibnamefont {Tung}},\
  }\bibfield  {title} {\enquote {\bibinfo {title} {{Natural emergence of
  cosmological constant and dark radiation from the
  Stephenson-Kilmister-Yang-Camenzind theory of gravity}},}\ }\href {\doibase
  10.1103/PhysRevD.88.123006} {\bibfield  {journal} {\bibinfo  {journal} {Phys.
  Rev. D}\ }\textbf {\bibinfo {volume} {88}},\ \bibinfo {pages} {123006}
  (\bibinfo {year} {2013})},\ \Eprint {http://arxiv.org/abs/1304.6334}
  {arXiv:1304.6334 [gr-qc]} \BibitemShut {NoStop}%
\bibitem [{Note4()}]{Note4}%
  \BibitemOpen
  \bibinfo {note} {We want to thank to M.~Blagojevi\'c, B.~Cvetkovi\'c and
  O.~Mi\v {s}kovi\'c for clarifications in this respect.}\BibitemShut {Stop}%
\bibitem [{\citenamefont {Gray}(1978)}]{gray1978einstein}%
  \BibitemOpen
  \bibfield  {author} {\bibinfo {author} {\bibfnamefont {Alfred}\ \bibnamefont
  {Gray}},\ }\bibfield  {title} {\enquote {\bibinfo {title} {Einstein-like
  manifolds which are not einstein},}\ }\href@noop {} {\bibfield  {journal}
  {\bibinfo  {journal} {Geometriae dedicata}\ }\textbf {\bibinfo {volume}
  {7}},\ \bibinfo {pages} {259} (\bibinfo {year} {1978})}\BibitemShut {NoStop}%
\bibitem [{\citenamefont
  {Derdzi{\'n}ski}(1980)}]{derdzinski1980classification}%
  \BibitemOpen
  \bibfield  {author} {\bibinfo {author} {\bibfnamefont {Andrzej}\ \bibnamefont
  {Derdzi{\'n}ski}},\ }\bibfield  {title} {\enquote {\bibinfo {title}
  {{Classification of certain compact Riemannian manifolds with harmonic
  curvature and non-parallel Ricci tensor}},}\ }\href@noop {} {\bibfield
  {journal} {\bibinfo  {journal} {Mathematische Zeitschrift}\ }\textbf
  {\bibinfo {volume} {172}},\ \bibinfo {pages} {273} (\bibinfo {year}
  {1980})}\BibitemShut {NoStop}%
\bibitem [{\citenamefont {Derdzi{\'n}ski}(1982)}]{derdzinski1982compact}%
  \BibitemOpen
  \bibfield  {author} {\bibinfo {author} {\bibfnamefont {Andrzej}\ \bibnamefont
  {Derdzi{\'n}ski}},\ }\bibfield  {title} {\enquote {\bibinfo {title} {On
  compact riemannian manifolds with harmonic curvature},}\ }\href@noop {}
  {\bibfield  {journal} {\bibinfo  {journal} {Mathematische Annalen}\ }\textbf
  {\bibinfo {volume} {259}},\ \bibinfo {pages} {145} (\bibinfo {year}
  {1982})}\BibitemShut {NoStop}%
\bibitem [{\citenamefont {Derdzi{\'n}ski}(1988)}]{derdzinski1988riemannian}%
  \BibitemOpen
  \bibfield  {author} {\bibinfo {author} {\bibfnamefont {Andrzej}\ \bibnamefont
  {Derdzi{\'n}ski}},\ }\bibfield  {title} {\enquote {\bibinfo {title}
  {Riemannian metrics with harmonic curvature on 2-sphere bundles over compact
  surfaces},}\ }\href@noop {} {\bibfield  {journal} {\bibinfo  {journal}
  {Bulletin de la Soci{\'e}t{\'e} Math{\'e}matique de France}\ }\textbf
  {\bibinfo {volume} {116}},\ \bibinfo {pages} {133--156} (\bibinfo {year}
  {1988})}\BibitemShut {NoStop}%
\bibitem [{\citenamefont {Cahen}\ \emph {et~al.}(2000)\citenamefont {Cahen},
  \citenamefont {Gutt},\ and\ \citenamefont {Rawnsley}}]{cahen2000symplectic}%
  \BibitemOpen
  \bibfield  {author} {\bibinfo {author} {\bibfnamefont {Michel}\ \bibnamefont
  {Cahen}}, \bibinfo {author} {\bibfnamefont {Simone}\ \bibnamefont {Gutt}}, \
  and\ \bibinfo {author} {\bibfnamefont {John}\ \bibnamefont {Rawnsley}},\
  }\bibfield  {title} {\enquote {\bibinfo {title} {Symplectic connections with
  parallel ricci tensor},}\ }\href@noop {} {\bibfield  {journal} {\bibinfo
  {journal} {Banach Center Publications}\ }\textbf {\bibinfo {volume} {51}},\
  \bibinfo {pages} {31} (\bibinfo {year} {2000})}\BibitemShut {NoStop}%
\bibitem [{\citenamefont {Jebsen}(1921)}]{Jebsen1921}%
  \BibitemOpen
  \bibfield  {author} {\bibinfo {author} {\bibfnamefont {J{\o}rg~Tofte}\
  \bibnamefont {Jebsen}},\ }\href@noop {} {\bibfield  {journal} {\bibinfo
  {journal} {Ark. Mat. Ast. Fys.}\ }\textbf {\bibinfo {volume} {15}} (\bibinfo
  {year} {1921})}\BibitemShut {NoStop}%
\bibitem [{\citenamefont {Birkhoff}(1923)}]{Birkhoff1923}%
  \BibitemOpen
  \bibfield  {author} {\bibinfo {author} {\bibfnamefont {Garrett~D.}\
  \bibnamefont {Birkhoff}},\ }\href@noop {} {\emph {\bibinfo {title}
  {Relativity and Modern Physics}}}\ (\bibinfo  {publisher} {Harvard University
  Press},\ \bibinfo {year} {1923})\BibitemShut {NoStop}%
\bibitem [{\citenamefont {Alexandrow}(1923)}]{Alexandrow1923}%
  \BibitemOpen
  \bibfield  {author} {\bibinfo {author} {\bibfnamefont {W.}~\bibnamefont
  {Alexandrow}},\ }\href@noop {} {\bibfield  {journal} {\bibinfo  {journal}
  {Ann. der Phys.}\ }\textbf {\bibinfo {volume} {72}},\ \bibinfo {pages} {141}
  (\bibinfo {year} {1923})}\BibitemShut {NoStop}%
\bibitem [{\citenamefont {Eisland}(1925)}]{Eisland1925}%
  \BibitemOpen
  \bibfield  {author} {\bibinfo {author} {\bibfnamefont {J.}~\bibnamefont
  {Eisland}},\ }\href@noop {} {\bibfield  {journal} {\bibinfo  {journal}
  {Trans. Amer. Math. Soc.}\ }\textbf {\bibinfo {volume} {23}},\ \bibinfo
  {pages} {213} (\bibinfo {year} {1925})}\BibitemShut {NoStop}%
\bibitem [{\citenamefont {Ramaswamy}\ and\ \citenamefont
  {Yasskin}(1979)}]{Ramaswamy:1979zz}%
  \BibitemOpen
  \bibfield  {author} {\bibinfo {author} {\bibfnamefont {Sriram}\ \bibnamefont
  {Ramaswamy}}\ and\ \bibinfo {author} {\bibfnamefont {Philip~B.}\ \bibnamefont
  {Yasskin}},\ }\bibfield  {title} {\enquote {\bibinfo {title} {{Birkhoff
  theorem for an $R+R^2$ theory of gravity with torsion}},}\ }\href {\doibase
  10.1103/PhysRevD.19.2264} {\bibfield  {journal} {\bibinfo  {journal} {Phys.
  Rev. D}\ }\textbf {\bibinfo {volume} {19}},\ \bibinfo {pages} {2264}
  (\bibinfo {year} {1979})}\BibitemShut {NoStop}%
\bibitem [{\citenamefont {Bergmann}\ \emph {et~al.}(1965)\citenamefont
  {Bergmann}, \citenamefont {Cahen},\ and\ \citenamefont
  {Komar}}]{bergmann1965spherically}%
  \BibitemOpen
  \bibfield  {author} {\bibinfo {author} {\bibfnamefont {P.~G.}\ \bibnamefont
  {Bergmann}}, \bibinfo {author} {\bibfnamefont {Michel}\ \bibnamefont
  {Cahen}}, \ and\ \bibinfo {author} {\bibfnamefont {A.~B.}\ \bibnamefont
  {Komar}},\ }\bibfield  {title} {\enquote {\bibinfo {title} {{S}pherically
  symmetric gravitational fields},}\ }\href@noop {} {\bibfield  {journal}
  {\bibinfo  {journal} {J. Math. Phys.}\ }\textbf {\bibinfo {volume} {6}},\
  \bibinfo {pages} {1--5} (\bibinfo {year} {1965})}\BibitemShut {NoStop}%
\bibitem [{\citenamefont {Hawking}\ and\ \citenamefont
  {Ellis}(1973)}]{hawking1973large}%
  \BibitemOpen
  \bibfield  {author} {\bibinfo {author} {\bibfnamefont {Stephen~W.}\
  \bibnamefont {Hawking}}\ and\ \bibinfo {author} {\bibfnamefont {George
  Francis~Rayner}\ \bibnamefont {Ellis}},\ }\href@noop {} {\emph {\bibinfo
  {title} {The large scale structure of space-time}}}\ (\bibinfo  {publisher}
  {Cambridge University Press},\ \bibinfo {year} {1973})\BibitemShut {NoStop}%
\bibitem [{\citenamefont {Peeters}(2007{\natexlab{a}})}]{Peeters2007550}%
  \BibitemOpen
  \bibfield  {author} {\bibinfo {author} {\bibfnamefont {Kasper}\ \bibnamefont
  {Peeters}},\ }\bibfield  {title} {\enquote {\bibinfo {title} {Cadabra: a
  field-theory motivated symbolic computer algebra system},}\ }\href {\doibase
  http://dx.doi.org/10.1016/j.cpc.2007.01.003} {\bibfield  {journal} {\bibinfo
  {journal} {Comput. Phys. Commun.}\ }\textbf {\bibinfo {volume} {176}},\
  \bibinfo {pages} {550} (\bibinfo {year} {2007}{\natexlab{a}})}\BibitemShut
  {NoStop}%
\bibitem [{\citenamefont {Peeters}(2007{\natexlab{b}})}]{peeters2007symbolic}%
  \BibitemOpen
  \bibfield  {author} {\bibinfo {author} {\bibfnamefont {Kasper}\ \bibnamefont
  {Peeters}},\ }\bibfield  {title} {\enquote {\bibinfo {title} {Symbolic field
  theory with cadabra},}\ }\href
  {http://www.fachgruppe-computeralgebra.de/CA-Rundbrief/car41.pdf} {\bibfield
  {journal} {\bibinfo  {journal} {Computeralgebra Rundbrief}\ }\textbf
  {\bibinfo {volume} {41}},\ \bibinfo {pages} {16} (\bibinfo {year}
  {2007}{\natexlab{b}})}\BibitemShut {NoStop}%
\bibitem [{\citenamefont {Peeters}(2007{\natexlab{c}})}]{Peeters:2007wn}%
  \BibitemOpen
  \bibfield  {author} {\bibinfo {author} {\bibfnamefont {Kasper}\ \bibnamefont
  {Peeters}},\ }\bibfield  {title} {\enquote {\bibinfo {title} {{Introducing
  Cadabra: A Symbolic computer algebra system for field theory problems}},}\
  }\href@noop {} {\  (\bibinfo {year} {2007}{\natexlab{c}})},\ \Eprint
  {http://arxiv.org/abs/hep-th/0701238} {arXiv:hep-th/0701238 [hep-th]}
  \BibitemShut {NoStop}%
\bibitem [{\citenamefont {Derdzi{\'n}ski}\ and\ \citenamefont
  {Shen}(1983)}]{Derdzinski01071983}%
  \BibitemOpen
  \bibfield  {author} {\bibinfo {author} {\bibfnamefont {Andrzej}\ \bibnamefont
  {Derdzi{\'n}ski}}\ and\ \bibinfo {author} {\bibfnamefont {Chun-Li}\
  \bibnamefont {Shen}},\ }\bibfield  {title} {\enquote {\bibinfo {title}
  {Codazzi tensor fields, curvature and pontryagin forms},}\ }\href {\doibase
  10.1112/plms/s3-47.1.15} {\bibfield  {journal} {\bibinfo  {journal}
  {Proceedings of the London Mathematical Society}\ }\textbf {\bibinfo {volume}
  {s3-47}},\ \bibinfo {pages} {15} (\bibinfo {year} {1983})}\BibitemShut
  {NoStop}%
\end{thebibliography}
\end{document}